\RequirePackage{fix-cm}
\documentclass{svjour3}                     % onecolumn (standard format)
\usepackage[a4paper,margin=3cm]{geometry}

\smartqed  % flush right qed marks, e.g. at end of proof
\usepackage{graphicx}
\usepackage{mathptmx}  
\usepackage{ltablex}
\usepackage{tabularx}
\usepackage{xcolor}
\usepackage{tablefootnote}
\usepackage{pdflscape}
\usepackage{amsmath}
\usepackage{amssymb}
\usepackage{caption}
\usepackage{subcaption}
\usepackage{graphicx}
\usepackage{adjustbox}
\usepackage[official]{eurosym}
\usepackage[flushleft]{threeparttable}
\usepackage{floatrow}
\usepackage{algorithm}
\usepackage[noend]{algpseudocode}
\usepackage{textcomp}
\usepackage{ulem}

\usepackage[
    backend=biber,
    style=apa,
    natbib
  ]{biblatex}
\appto{\bibsetup}{\raggedright}
\usepackage{hyperref}
\hypersetup{
unicode=true,                 % non-Latin characters in Acrobat bookmarks
pdftoolbar=true,              % show Acrobat toolbar?
pdfmenubar=true,              % show Acrobat menu?
pdffitwindow=true,            % page fit to window when opened
pdfnewwindow=true,            % links in new window
colorlinks=true,              % false: boxed links; true: colored links
linkcolor=black,                % color of internal links
citecolor=black,                % color of links to bibliography
filecolor=blue,               % color of file links
urlcolor=black,                % color of external links
bookmarksopen=true,
}
\newfloatcommand{capbtabbox}{table}[][\FBwidth]
\journalname{Journal of Real Estate Finance and Economics}
\begin{document}

\title{Irish Property Price Estimation Using A Flexible Geo-spatial Smoothing Approach: What is the Impact of an Address?}


\titlerunning{What is the Impact of an Address?}        % Short form of title if too long for running head

\author{Aoife K. Hurley         \and
        James Sweeney 
}


\institute{A.K. Hurley \at
              Department of Mathematics and Statistics, 
              University of Limerick, Limerick, Ireland V94 T9PX\\
              \email{Aoife.Hurley@ul.ie}           %  \\
           \and
           J. Sweeney \at
              Department of Mathematics and Statistics, 
              University of Limerick, Limerick, Ireland V94 T9PX \\
}

\date{}

\maketitle

\begin{abstract} % 150 to 250 words

Accurate and efficient valuation of property is of utmost importance in a variety of settings, such as when securing mortgage finance to purchase a property, or where residential property taxes are set as a percentage of a property's resale value. Internationally, resale based property taxes are most common due to ease of implementation and the difficulty of establishing site values. In an Irish context, property valuations are currently based on comparison to recently sold neighbouring properties, however, this approach is limited by low property turnover. National property taxes based on property value, as opposed to site value, also act as a disincentive to improvement works due to the ensuing increased tax burden. In this article we develop a spatial hedonic regression model to separate the spatial and non-spatial contributions of property features to resale value. We mitigate the issue of low property turnover through geographic correlation, borrowing information across multiple property types and finishes. We investigate the impact of address mislabelling on predictive performance, where vendors erroneously supply a more affluent postcode, and evaluate the contribution of improvement works to increased values. Our flexible geo-spatial model outperforms all competitors across a number of different evaluation metrics, including the accuracy of both price prediction and associated  uncertainty intervals. While our models are applied in an Irish context, the ability to accurately value properties in markets with low property turnover and to quantify the value contributions of specific property features has widespread application. The ability to separate spatial and non-spatial contributions to a property's value also provides an avenue to site-value based property taxes. 

% 4 to 6 Keywords
\keywords{House Prices \and Automatic Valuation Models \and Spatial Models \and Generalised Additive Models \and Hedonic Regression \and Non-linear Smoothing}
\end{abstract}

\section{Introduction}
\label{intro}
There is substantial public and commercial interest in accurate and efficient automated approaches to property valuation. The purchase of a property is typically the largest lifetime financial outlay for most households and so there is an intrinsic interest in accurate estimation of the value of owned, or prospectively owned, property. From a debt financing perspective, property valuations are typically required by banking institutions to secure mortgage finance prior to property purchase, or when refinancing an existing loan. Similarly, an investment fund, bank or government agency may require regular valuations of assets under their ownership for financial reporting reasons. The price of  a  property  will depend on a wide variety of factors, ranging from property specific features such as its type, size, quality of finish, in addition to its location in terms of proximity to schools and public transport services. However, the contribution and importance of each of these individual aspects to the value of a property will vary from country to country depending on national preferences.

The valuation of properties is also of substantial governmental interest in terms of revenue generation via property taxes. In Ireland, as in a number of countries, property taxes are based on the value of the property itself, as opposed to a tax on its location or site value (\citet{OECDProperty}).  This method of taxation is primarily due to the ease in estimating property resale values. It can, however, act as a deterrent in undertaking improvement works that increase property values, due to the resulting increased tax. This has wider societal impacts in areas such as climate, with energy inefficient homes substantially contributing to greenhouse gas emissions - the residential sector accounted for 27\% of all energy usage in Ireland in 2020. %(https://www.gov.ie/en/publication/a4d69-long-term-renovation-strategy/). 
There is also an incentive to mitigate the tax penalty through under-utilisation by letting a property fall into disrepair and thus having a lower value. On the other hand, a site value tax (SVT), which does not penalise improvement works, requires the separation of a property's value into the constituent parts of building value and the value of the underlying site. This site value is derived from its location and access to services, facilities and utilities. Moves towards a full site value tax (SVT) are impeded by the difficulty in valuing land - only 3 OECD members (Sweden, Estonia and Australia) have a tax of this form (\citet{OECDProperty}).  

Automated valuation models (AVMs) provide an efficient approach to large scale property valuations, being based on statistical models, with price prediction models constructed for several major cities worldwide. They have substantial potential in the context of property tax calculations for property owners, due to their providing objective property valuations. AVMs can be broadly separated into two approaches at present, comprising approaches focused on either hedonic geospatial regression modelling approaches, or alternatively taking a machine learning perspective. The primary benefit of a spatial hedonic modelling approach is the benefit of insights into the contribution of respective property features to price. \citet{GELFAND_Singapore} provide a geo-spatial model with repeat sale measures for price prediction of condominiums in Singapore. \citet{Liu_2013} investigate the property market of the Dutch Randstad region with property variables including property type, age and size. \citet{Oust:2020aa} account for repeat sales as well as some property specific attributes including floor area, construction year and garden access in an application to the Oslo housing market.
 In each case, the authors have access to large data sets ranging from 16,000 to 438,000 transactions. In contrast from an explanation perspective, machine learning or artificial intelligence based approaches to property price prediction are typically more focused on predictive error as opposed to explainability of models. The drawback for many of the more complex approaches, such as neural networks, is the difficulty in interpreting the relationship between selling price and house features, in addition to the volume of data required to accurately fit models. \citet{ML_WHO} apply support vector regression, random forest and gradient boosting machine to a Hong Kong data set containing 39,554 housing transactions. Their best model has an $R^2$ of $90\%$ though they report high bias in the model predictions for extreme values. \citet{phan2018housing} evaluates the performance of various machine learning algorithms, including regression trees and neural networks to the Melbourne property market noting difficulties in the interpretation of the prediction output and over-fitting issues. In general, machine learning approaches require substantial data sets to train models, with overfitting an issue in smaller ones. It is not clear that the methods would have similar accuracy results in areas with much lower property turnover.
 
 In this article, we focus on the development of geo-spatial statistical models that have improved price prediction performance and interpretation for small data sets in regions where property transactions are low or relatively infrequent. We explore the impact of address mislabelling and allow for public bias in postcode allocation in the Dublin market, where property addresses are misreported. Our focus is on explaining the contribution of each property component to overall price - we evaluate the impact of improvement measures such as improved energy efficiency or renovation, on the value of a property, allowing for a deeper understanding of the contribution and dominance of property specific aspects on value.  Finally, we  provide a speculative framework using estimated land values for the development of a land value based tax system in an Irish context which may allow for a more equitable distribution of property taxes. A key benefit of our model-based approach is the accounting for uncertainty in model parameters in ensuing price prediction intervals - machine learning algorithms by default usually provide point estimates only, and so decisions are made ignoring the uncertainty surrounding these estimates.

 Our article is organised as follows. In Section~\ref{sec:Data} we introduce our data set for the Dublin property market and provide an exploratory analysis of the links between spatial location and property features with price. In Section~\ref{sec:methodology} we outline our statistical modelling methodology and we outline our models and selections required for a Gaussian Process smooth. Section~\ref{sec:Results} presents the results, while the final section concludes the article.

%%%%%%%%%%%%%%%%%%%%%%%%%%%%%%%%%%%%%%%%%%%%%%%

\section{The Irish Property Market}
\label{sec:Data}

Irish people have historically had a strong preference for home ownership over renting, with a home ownership rate of 70.3\% in 2018 (\citet{Eurostat2018}). In most cases, properties are sold by private treaty. Bidders submit their offers to the selling agent who liaises with the vendor, with the property typically sold to the highest bidder. On average only $2\%$ of properties change hands per year, with the average house transacted approximately once every 60 years (\citet{Maguire2016}). For property valuations, property agents base estimates on a combination of their opinion and prices of \textit{similar} nearby sold properties. This approach suffers from a number of deficits. First, due to low property turnover, the number of nearby comparison properties is typically low. The comparison properties may also be of a completely different type, for example apartments as opposed to houses. Second, the Irish residential property price register only includes date of sale, price and address, and does not provide property specific information. Consequently, there can be a large degree of variation in estimates from a nearest neighbours type approach. Due to these issues, there has been little to no progress in the development of automated property valuation models (AVMs) for Ireland.

\begin{figure}[h!]
    \centering
    \includegraphics[width = \textwidth]{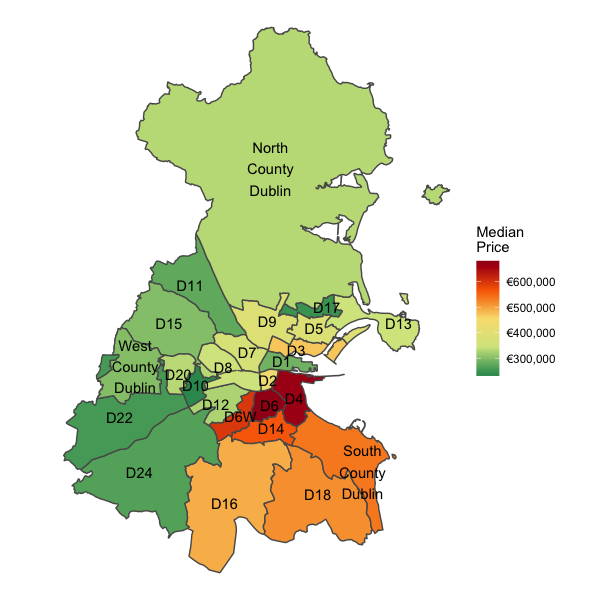}
    \caption{Median price per postcode}
    \label{fig:medppcode}
\end{figure}

Within Ireland, the Dublin property market accounts for approximately one third of all property transactions and provides the focus for this article. 
The county of Dublin has 25 postcode regions shown in Figure~\ref{fig:medppcode}, with a breakdown of the median property price per postcode region. For clarity, postcodes have been shortened from Dublin X to DX. The river Liffey bisects the city into north and south - for numbered postcodes, the even numbers are south of the river and odd numbers are to the north.  Though a generalisation, postcodes in the south-east of the city and south county are perceived to be the more affluent areas with the north inner city areas seen to be more economically deprived (\citet{DeprivationinCity}). There has long been speculation that people would pay more to have a perceived affluent postcode on their address. 

\subsection{The Dublin House Price data set}

Properties that are for sale are typically listed on at least one of a number of online property portals, with \textit{Daft.ie} and \textit{MyHome.ie} the largest providers on the Irish market. The data set available for exploration and model development was provided by 4Property Ltd, containing  property sales in Dublin (both city and county) for dates between January 2018 and November 2018 inclusive, amounting to 5,285 transactions in total. This was reduced to 5,208 properties after removing properties with incorrect information such as size, or with floor areas below minimum size thresholds of 37m$^2$. In Table~\ref{tab:dataset} we have listed the data fields provided with each property. The `Description' variable includes detailed extensive descriptive information on individual properties that is, to the best of our knowledge, not available to historical researchers, particularly in an Irish context. At the property level, the description typically includes details on features such as property aspect, heating systems used and general property condition. In Section~\ref{sec:MiningVars} we outline our approach for mining the description for key words speculated to be influential in previously published research or via expert opinion on influential property variables. Unlike other major cities worldwide, age does not play a major role when advertising properties in Dublin. Parts of the city were developed at different times and so building stock and its age may be loosely associated with postcodes that have been developed at different times. Energy rating is sometimes also a proxy for age with newer dwellings having much higher energy ratings than historic stock. 

\begin{table}[h!]
    \centering
    \caption{Description of Data Set}\label{tab:dataset}
    \begin{tabular}{ l  l  }
        \hline
        \textbf{Variable} & \textbf{Description} \\
        \hline
        Price & Sale price of property in € \\
        Sale Date & Date the property was sold \\
        Latitude & Latitude co-ordinate of the property \\
        Longitude & Longitude co-ordinate of the property \\
        Neighbourhood & Townland, village or town in which the property \\ 
         & is located\\
        Baths & Number of baths in the property \\
        Beds & Number of beds in the property \\
        BER & Building Energy Rating (BER) for the property,  properties \\
         & less than $50 m^2$ are exempt - are not required to have a rating \\
        Description & A written description that was used \\
         & in the advertising of the property \\
        Size & The floor area of the property in $m^2$  \\
        Property Type &  Classification of property type, \\
         & e.g. Duplex, End of Terrace House, Semi-Detached House, etc. \\
        Postcode & Original postcode provided by the vendor for the property \\
        \hline
    \end{tabular}
\end{table}

\begin{figure}[h!]
    \centering
    \includegraphics[width=\textwidth, height = 11cm]{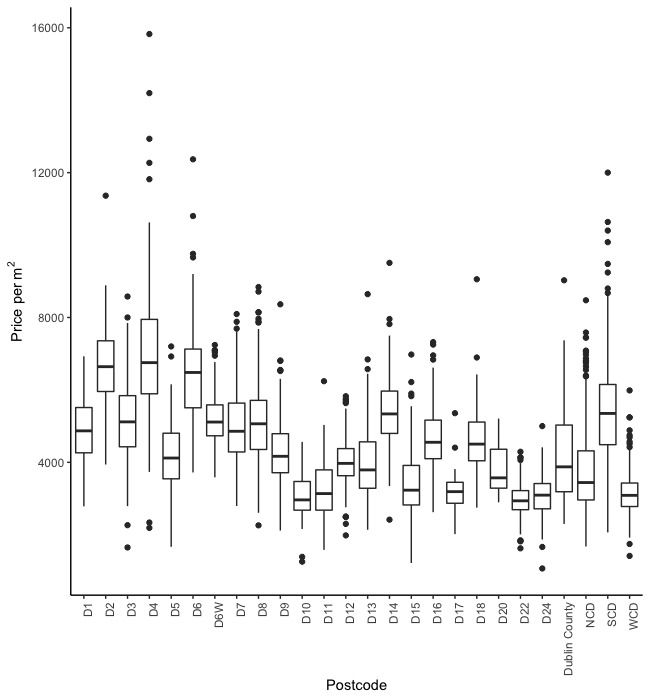}
    \caption{Variation in sale price per $m^2$}
    \label{fig:bp_pm2p}
\end{figure}

The mean sale price for the 5,208 properties sold between January 2018 and November 2018 is €484,443, with the median price of €392,000, highlighting the skewed nature of property prices in the data set. We alleviate this issue by standardising price by floor area, resulting in a mean and median prices per $m^2$ of €4,569 and €4,405 respectively. There is little variation in size across the 25 postcodes. In Figure~\ref{fig:bp_pm2p} we provide a breakdown of the price per $m^2$ by postcode. Within postcodes, the distribution of price per m$^2$ is broadly symmetric. We notice a higher degree of variability in postcodes that have span large areas and have a diverse range of people of differing socioeconomic backgrounds. One example of increased variability is seen in Dublin 4 which encompasses a large area with various levels of affluence, and subsequently has a broad interquartile range, as well as having multiple outliers. This is unlike Dublin 20, which covers a small area with little variation in socioeconomic backgrounds and this is seen in the variation in price per m$^2$. 
The ordering of the postcodes is both alphabetical and can be coarsely equated to distance from the city centre. We would expect price per $m^2$ to reduce with increased distance, however, D14 and D16 have higher medians as compared to postcodes of a comparable distance from the centre of the city. This could be down to many reasons including the light rail system (LUAS) servicing these postcodes providing a sought-after link with the city centre. Alternatively the impact could be for socioeconomic reasons with these postcodes having an increased valuation for perceptions of affluence.

In Table~\ref{tab:Postcodes} we provide an overview of the number of properties and pricing information for each property type and within each postcode. For some postcodes, such as Dublin 10 and Dublin 17, we witness extremely low property turnover, which would pose problems for nearest neighbour based methods. The table also provides details of the breakdown of properties by type. We observe the dominance of certain property types within postcodes, e.g. D1 and apartments, and others have a broad mix, e.g. North County Dublin. When there is a dominant property within a postcode, valuations of other property types within that postcode is difficult. 

\begin{figure}[h!]
    \centering
    \includegraphics[width = \textwidth]{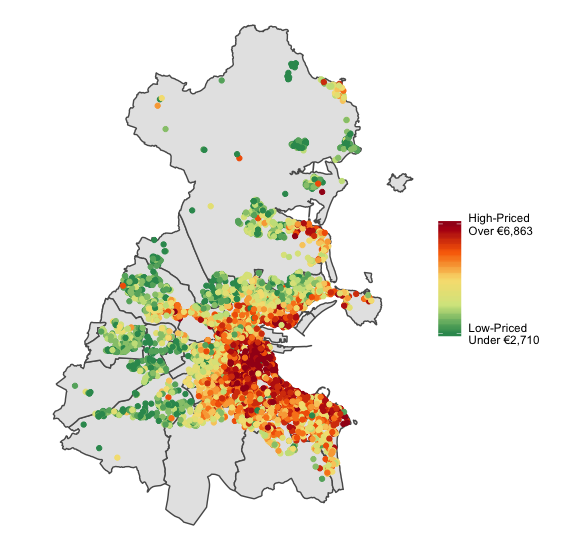}
    \caption{Price per $m^2$}
    \label{fig:PPM_Map}
\end{figure}

In Figure~\ref{fig:medppcode} we present the median price of properties within each postcode, averaged over dwelling type and size. We observe a spatial pattern that has higher median prices clustered together in the south city along the coast and lower median prices along the north and west county border. 
In Figure~\ref{fig:PPM_Map} we visualise the spatial structure in price per $m^2$ across Dublin. To do so, we create 14 groups of approximately equal size and group based on observed price per $m^2$ values. These groupings make clear the variation in prices across the county of Dublin and is robust to outliers in the data. Similar to Figure~\ref{fig:medppcode}, higher values (dark red) are concentrated together around the city centre, to the south of the city and along the coast. Lower values, shown in dark green, are observed as one moves away from the city centre towards the county border.

\begin{table}[h!]
    \caption{Number of Properties and Median Price per $m^2$ per Postcode}
    \label{tab:Postcodes}  
    \begin{threeparttable}
    \centering
    \resizebox{\textwidth}{!}{\begin{tabular}{ l c c c c c c  c c c }
        \hline
        \textbf{Postcode} & \textbf{Apartment} & \textbf{Detached} & \textbf{Duplex} & \textbf{End of } & \textbf{Semi-Detached} & \textbf{Terraced } & \textbf{Townhouse} & \textbf{Number of} & \textbf{Median Price} \\
         & & \textbf{House} & & \textbf{Terrace House} & \textbf{House} & \textbf{House} & & \textbf{Properties} & \textbf{ per $m^2$}\\
        \hline
        D1                      &  51 &   0 &   1 &   2 &   0 &   7 &   0 & 61  & €4,869 \\
        D2                      &  45 &   0 &   0 &   1 &   1 &   5 &   2 & 54  & €6,641 \\
        D3                      &  27 &   9 &   0 &  21 &  33 &  62 &   0 & 152 & €5,117 \\
        D4                      &  72 &  22 &   3 &  20 &  38 &  73 &   8 & 236 & €6,753 \\
        D5                      &  10 &  12 &   3 &  26 &  82 &  59 &   0 & 192 & €4,118 \\
        D6                      &  39 &  15 &   3 &  19 &  36 &  60 &   5 & 177 & €6,481 \\
        D6W                     &  17 &  11 &   1 &  10 &  57 &  22 &   0 & 118	& €5,112 \\
        D7                      &  38 &   7 &   4 &  32 &  23 & 107 &   0 & 211 & €4,857 \\
        D8                      & 117 &   1 &   5 &  26 &   7 & 112 &   4 & 272 & €5,064 \\
        D9                      &  53 &   9 &   3 &  23 & 101 &  59 &   2 & 250 & €4,166 \\
        D10                     &   5 &   2 &   0 &   9 &   0 &  14 &   0 & 30  & €2,962 \\
        D11                     &  28 &   4 &   2 &  15 &  35 &  39 &   2 & 125 & €3,136 \\
        D12                     &   7 &  10 &   0 &  45 &  35 &  86 &   0 & 183 & €3,971 \\
        D13                     &  24 &  15 &   6 &  15 &  45 &  30 &   0 & 135 & €3,790 \\
        D14                     &  26 &  25 &   1 &  12 &  93 &  39 &   4 & 200 & €5,337 \\
        D15                     &  74 &  46 &  22 &  20 & 117 &  32 &   2 & 313 & €3,231 \\
        D16                     &  32 &  19 &   1 &   6 & 115 &   7 &   2 & 182 & €4,553 \\
        D17                     &  13 &   3 &   1 &   2 &   8 &   4 &   0 & 31  & €3,190 \\
        D18                     &  76 &  60 &   4 &   9 &  59 &  14 &   4 & 226 & €4,503 \\
        D20                     &   5 &   1 &   1 &   1 &   8 &   8 &   0 & 24  & €3,571 \\
        D22                     &  14 &   7 &   2 &   7 &  30 &  18 &   0 & 78  & €2,937 \\
        D24                     &  34 &   7 &   4 &  17 &  37 &  28 &   2 & 129 & €3,093 \\
        NCD                     & 145 & 107 &  34 &  63 & 176 & 129 &   7 & 661 & €3,441 \\
        SCD                     & 209 & 140 &  21 &  42 & 267 & 128 &   8 & 815 & €5,350 \\
        WCD                     &  58 &  21 &  12 &  18 & 132 &  36 &   5 & 282 & €3,088\\
        Dublin County  &  31 &  14 &   5 &   5 &   8 &   8 &   0 & 71  & €3,875 \\
        \textbf{Total}          & \textbf{1,250} & \textbf{567} & \textbf{139} & \textbf{466} & \textbf{1,543} & \textbf{1,186} & \textbf{57} & \textbf{5,208} & \textbf{€4,405}\\
        Median Price per $m^2$ &  €4,573 & €4,743 & €3,096 & €4,150 &  €4,320 & €4,389 & €5,078 & & \\
        \hline
    \end{tabular}}
    \end{threeparttable}
\end{table}

\subsection{Text Mining of Price Prediction Features}\label{sec:MiningVars}
Valuable information is contained in the textual summary of a property, requiring the use of natural language processing tools to extract features of interest. There is some published research identifying such features: \citet{GardenEffect} and \citet{Mayor2009hedonic} explore the impact of gardens on price with an increase in explanatory performance of their models and reduced prediction error.
 \citet{LU201824} found that dwelling units with southerly aspects have higher property value in an examination of the Shanghai housing market. \citet{Liverpool} noted the improved predictive performance of models for the Liverpool housing market with the inclusion of garage, garden and central heating. \citet{culdesac} ventures that there is a premium for having a property on a cul de sac.  \citet{KIEFER2011249} include a variable for fireplace in their models and notes the positive impact on price. \citet{Oust:2020aa} include the variables `penthouse', and `garden' amongst others in their models. Several articles, including \citet{VANOMMEREN201125}, investigate the impacts of parking policy on house prices. \citet{Wittowsky_2020} construct an ordinary least squares (hedonic) model and a spatial lag model to investigate the impact of different attributes including accessibility, `walk score' and socioeconomic characteristic of neighbours on residential housing prices. However, the lack of demographic information at the household level in an Irish context means that such variables cannot be included.

We performed a simple, case insensitive string search for certain housing characteristics with the descriptive phrases identified driven by previous literature as well as expert opinion on features of specific interest in an Irish context. The plot or site size is an obvious important variable in the valuation of a property, however this information is not explicitly available with each transaction listing. Only 172 properties (3\% of data) mentioned ``acre" in their description. In addition, the average plot size in Dublin varies by postcode and is not easily obtained. We attempt to identify a proxy for this variable by identifying properties where development potential or large gardens are prominent within textual descriptions. In Table~\ref{tab:scrapDesc} we list the words and phrases mined within each textual description in addition to their numbers of occurrence. In addition to these variables, `Attic Conversion', `Development Potential', `Open Plan',  `Immersion' and `Hot Press' were identified. 
Converting attics into another living space to increase available floor area is becoming more common in the Irish property market due to the reduced cost in comparison to an increase in property footprint. A property with development potential for an extension would generally sell for more with the opportunity for the buyer to increase property size or build additional units on the site. Open plan living spaces are also extremely popular at present. `Immersion' is a reference to water heating being electric and thus more expensive typically. Due to the proportion of rainy days in Ireland access to an airing closet is preferred for clothes drying, resulting in the inclusion of `hot press'.

\begin{table}[h!]
    \caption{Attributes Extracted from `Description' variable}
    \label{tab:scrapDesc}    
    \centering
    \begin{tabular}{l  c }
        \hline
        \textbf{Word or Phrase(s)} & \textbf{Number of Observations} \\
        \hline
        South Facing or South Orientation & 871\\
        Attic Conversion & 190\\
        Parking or Car Spot/Space & 3782\\
        Development Potential & 492\\
        Open Plan & 1512\\
        Fireplace & 2403\\
        Central Heating & 2596\\
        Immersion & 196\\
        Hot Press & 993\\
        Garage & 770\\
        Large Garden & 155\\
        Ground Floor Apartment & 201 \\
        First Floor Apartment & 119 \\
        Second Floor Apartment & 66 \\
        Penthouse Apartment & 39\\
        Refurbished/Renovated/Upgraded & 1219\\
        Cul-de-Sac & 1006\\
        \hline
    \end{tabular}
\end{table}

Most hedonic regression models contain some distance measures as a proxy for spatial location, such as to the Central Business District (CBD). We construct similar variables with distance measures calculated as the shortest distance between two points on an ellipsoid. We assign the International Financial Services Centre, or IFSC, as Dublin's CBD. We also construct a number of categorical indicator distance variables with the distance tapered to have zero affect above a distance threshold. These include indicator variables for proximity to parks, transport hubs and the city centre - the distance measures taken and the radius distance are presented in Table \ref{tab:dists}. Additionally, interactions terms for `Garden' and `Parking' within city centre areas were taken. 
This is due to the high demand for parking and gardens within the city centre for a 2km radius, which we believe would add value and reduce modelling errors. Properties outside the city centre typically have free parking and garden access and so these aspects would not add to their value.

\begin{table}[h!]
    \centering
    \caption{Distance Measures}
    \label{tab:dists}
    \begin{tabular}{l  l l}
    \hline
        \textbf{Location} & \textbf{Maximum Distance} & \textbf{Data Type} \\
    \hline    
        International Financial Services Centre (IFSC) & NA & Continuous \\
        Dublin Airport & 5KM & Categorical\\
        O'Connell Bridge (City Centre) & 2KM & Categorical \\
        Dublin Area Rapid Transit (DART) Station & 1.5KM & Categorical \\
        LUAS (Light Rail) Stop & 1KM & Categorical \\
        Park or Public Green & 5KM & Categorical \\
    \hline
\end{tabular}

\end{table}

%%%%%%%%%%%%%%%%%%%%%%%%%%%%%%%%%%%%%%%%%%%%%%%

\section{Automated Valuation Models} \label{sec:methodology}

In more recent years, machine learning approaches such as K-Nearest Neighbours, Decision Trees and Random Forests have been applied in real estate research. For some of these methods, interpretability can be an issue, as well as requiring large data sets to accurately fit the models. From a statistical modelling standpoint, hedonic regression, generalised additive models (GAMs), regression kriging and geographically weighted regression (GWR) have also been used. These approaches have lower data requirements for parameter estimation and typically have greater interpretability. 

In this section we introduce the suite of models we apply to the data set in order of complexity including; classic  hedonic regression, a nearest-neighbours based approach and geo-spatial based models. 

\subsection{Hedonic Regression}

Forms of hedonic regression are used throughout the literature for property price prediction, generally as a baseline model as a comparison to recent modelling advances. 
\citet{Oust:2020aa} use hedonic models that have additional terms that combine single and repeat sale measures and compare these to their geographically weighted regression approach. 
\citet{LU201824} construct a hedonic pricing model for the Shanghai housing market in order to estimate the impact of property aspect. Similarly, \citet{Mayor2009hedonic} use a hedonic house price model to estimate the effects of green spaces on property prices. 

Our hedonic regression model takes the form

\begin{equation}
    \mathrm{log}_e(\boldsymbol{V_i}) = \boldsymbol{\beta_0} + \sum_k X_{k i} \boldsymbol{\beta_{k}} 
    + \sum_l Z_{l i} \boldsymbol{\gamma_{l}}
    +
    \boldsymbol{\epsilon_i}    
\end{equation} \label{eq1}

\noindent where $\boldsymbol{V_i}$ represents the price per $m^2$ of property $i$ and $\epsilon_i \sim \mathcal{N}(0,\sigma^2)$ is iid unstructured Gaussian noise. It is typical to model log price per $m^2$, being less affected by the skewness introduced by the right tailed nature of house prices in comparison to a lower price bound at 0. Our model is specified as follows: $\boldsymbol{\beta_0}$ is an intercept term relating to the mean log(price) per $m^2$. We group the property characteristics into continuous effects $X_i$ = $\{\mathrm{size}_i,\mathrm{beds}_i,
\mathrm{baths}_i,\mathrm{CBD}_i\}$, and dummy (or intercept) indicator labels $Z_i$ comprised of the features outlined in Table~\ref{tab:scrapDesc} in addition to property postcode, property type and energy rating. $\boldsymbol{\beta_k}$ and $\boldsymbol{\gamma_l}$ represent the regression scaling coefficients associated with variables $\boldsymbol{X_k}$ and $\boldsymbol{Z_l}$ respectively. 
These models are easy to fit using standard least squares or maximum likelihood methods, and their properties are well known. Spatial aspects in the data set are coarsely modelled through the postcode indicator variables and the distance measures for proximity to parks and transport services. As we will observe in later sections, this model is not sufficiently complex to account for non-linear structured patterns in some of the model variables as well as the distance measures.

\subsection{Machine Learning Based Approaches}

There is increasing interest in the application of machine learning and artificial intelligence approaches to price prediction in automated valuation models with a number of commercial operators, such as Zillow\textsuperscript{\textregistered}  offering services in this area. 
However, machine learning approaches typically require vast volumes of data to accurately fit models, with this issue exacerbated in situations where there are complex relationships between predictor variables and the target of interest. A further drawback is that the interpretability of these methods can dwindle where more complex machine learning approaches are utilised. Decision Trees and Regression Trees (RT) are simple to understand given the partitioning of the feature space into rectangles (\citet{hastie2009elements}), however complex approaches such as neural networks are much less interpretable. Studies into the interpretability of neural networks are ongoing (\citet{NN_Interpret_Survey}, \citet{fan2021interpretability}).

\citet{ML_WHO} apply multiple machine learning approaches, including random forest and gradient boosting, to their data set for Hong Kong containing 39,554 housing transactions from June 1996 through to August 2014. They reported that although their best model has an $R^2$ of $90\%$, there is a high bias in the model predictions for extreme values. \citet{ML_MarketAnalysis} apply the eXtreme Gradient Boosting (XGB) algorithm for rental prediction on a dataset for over 52,000 apartments in Frankfurt am Main, Germany. On these predictions, Interpretable Machine Learning (IML) methods are applied to examine both feature importance and effects. From these IML methods, they find that living area, age and distance measures to the CBD and department stores are the most influential for rental prediction. \citet{PaceHayunga2020} examine over 80,000 observations from the Dallas multiple listing service, fit a traditional OLS model and to these results fit a CART (classification and regression trees) model. They found the residuals of the OLS model contained further information that the CART approach could explain. 
\citet{Oust:2020aa} use a nearest-neighbours approach examining up to 120 nearest neighbours to correct the residuals of a hedonic regression model, although their approach can be considered a geo-spatial one as they use geographically weighted regression to account for spatial correlation. 
\citet{phan2018housing} evaluates various machine learning algorithms, including regression trees and neural networks, to predict property prices. The data set used is for Melbourne, and the author notes difficulties in the interpretation of the prediction output of neural network as well as over-fitting issues in some of the methods used.
\citet{PARK20152928} implement a number of machine learning algorithms, including decision trees, to examine the relationship between property listing and closing prices. They show that all methods implemented worked well, but note they only investigate one type of property in one location.

\paragraph{K-nearest neighbours}
The most prevalent approach to valuation in an Irish context is a form of k-nearest neighbours averaging, as outlined in Algorithm~\ref{alg::nns}. In order to control for the effect of property type, we group comparison properties by type to ensure that the neighbours identified for each property are the same. We also control for the impact of property size by basing estimates on the price per $m^2$ of neighbouring properties as opposed to raw price alone.  

\begin{algorithm}
\caption{Nearest Neighbours Approach}\label{alg::nns}
\begin{algorithmic}[1]
    \For {$i=1,\dots, 5208$}

    \For {$k= 3, 5, 7, 9$}
        \State Create subset of data set of type property $i$, $Y^{(i)}$.  
        \State Identify set $\{y_1^{(i)},\ldots,y_k^{(i)}\}$  of the $k$ spatially nearest neighbours to property $i$
        \State Estimate price per $m^2$ as $\Tilde{y}_i$ = median$\{y_1^{(i)},\ldots,y_k^{(i)}\}$
    \EndFor
     \State $\widetilde{\mathbf{Y}}_{i}$ 
    \EndFor
    \Return{K nearest neighbour estimates $\{\widetilde{\mathbf{Y}}_{i}, 1, \ldots, 5208\}$}
\end{algorithmic}
\end{algorithm}

The major strength of this approach is the ability to incorporate spatial correlation in predictions in terms of location and postcode by utilising neighbouring properties. We also control for two of the primary value driving features in property type and size. However, the accuracy of this approach is greatly affected by a lack of neighbouring comparable properties, in addition to not controlling for the characteristics of these properties. As outlined in Table~\ref{tab:Postcodes}, the property types in a locality can be quite heterogeneous in nature and we observe the impact of this in the accuracy of price predictions using this method in Section~\ref{sec:Results}. 

\paragraph{More complex methods}
Decision trees are a non-parametric regression approach, which can be applied to both classification (Classification Tree) or regression (Regression Tree) problems. The tree contains a set of sequential control statements, which once completed leads to a terminal node or decision. These control statements partition the feature space (space spanned by predictor variables) into a set of rectangles (\citet{strobl2009introduction}). These partitions group similar observations together, and once the partitioning is completed a constant value is predicted within each area. For a regression tree (RT), the tree starts with an observation and concludes with a target value for the variable we wish to predict. One drawback of RTs is that a large tree may overfit the data whereas a tree that is too small may miss important structures. Trees are also susceptible to changes in the data, as these can lead to different splits along the tree, resulting in high variances (\citet{hastie2009elements}).

Random forests (\citet{breiman_randomforests}) take a large collection of de-correlated decision trees and averages them to reduce the variance of the estimated prediction function. In generating the random forest, the algorithm samples both the observations and variables. Since the random forest determines automatically which variables are important, it is possible to investigate variable importance. For regression problems, the output value is that of the predictions from each tree in the forest averaged. Random forests typically outperform decision trees but are not as accurate as gradient boosted trees. Unlike decision trees, random forests are considered to be ``black box". This is due to the intuitive rules of the decision tree being lost (\citet{bruce2017practical}). 

Boosting (\citet{freund1997decision}) is a technique used to create an ensemble of models, and is commonly used with decision trees. Each successive model fitted tries to minimise the error of the previous model. With each iteration, the weak rules (performs just  slightly  better  than  random  guessing) from each individual classifier are combined to form one, strong prediction rule. There are several variants to the boosting algorithm originally proposed including  AdaBoost(\citet{freund1997decision}) and gradient boosting. Gradient boosting is set up to minimise a cost function and uses the gradient to determine how to tune the model parameters.

\citet{hastie2009elements} compare random forests and gradient boosting on a California housing data set. They find when examining the mean absolute error, the gradient boosting method outperforms the random forests. They also note that in this example, the mean absolute error stabilises for the random forests much sooner than that of the gradient boosting and the weaker boosting model outperforms the random forests.

In Section~\ref{sec:MLResults} we discuss the performances of decision tree and random forest approaches to model fit. In an Irish context, the primary issue with applying these models is the lack of sufficient data to train models, particularly outside of large urban areas.

\subsection{Geo-spatial Modelling}

The primary benefit of a geo-spatial approach to property price prediction is that it enables the reduction of uncertainty in areas with low numbers of property transactions. This is achieved by leveraging information from neighbouring areas where there is more data to give more certainty in predictions for areas with less data.  \citet{Basu_1998} examine the spatial autocorrelation in a  data set of over 5000 transactions of single-family homes in Dallas, Texas. They propose that adjacent properties might have similar observable and unobservable characteristics resulting in properties in close proximity being similarly priced. \citet{Farber_2006} compare four models on 19,007 housing sales in Canada. These comprise: an ordinary least squares model, a spatial autoregressive regression model, a geographically weighted regression (GWR) model and a moving window regression model. They report the GWR model as obtaining the highest $R^2$, but the authors do note some disadvantages to GWR including irrational coefficients. \citet{Gao_2006} propose an alternative way of empirically evaluating models, using house and land price data set in Tokyo from \citet{Gao_2001} concluding that if a spatial relationship is implied in the data, spatial models outperform a simpler model. \citet{Gelfand_2004} use a Gaussian process based approach to model spatial correlation in a Louisiana housing data set noting that spatial effects emerge as very important in explaining house prices. \citet{Bourassa_2007} implement both Conditional Autoregressive (CAR) and Simultaneous Autoregressive (SAR) models, as well as other geostatistical models, to a data set containing 4,800 transactions for Auckland, New Zealand. They find that the CAR and SAR methods perform poorly when compared to the other geo-statistical models they fit.  \citet{Liu_2013} compare the prediction performance of a spatiotemporal autoregressive model to a traditional hedonic model. The analysis is performed on a comprehensive housing transaction data set from the Dutch Randstad region, which  spans ten years, 1997 to 2007 with 437,734 transactions. The data set contains variables on both structural attributes and the region or submarket the house is located. This article found that spatial dependence was of a larger magnitude than that of temporal dependence. Liu concluded that accounting for spatial and temporal dependence is not only theoretically warranted, but contributes to better prediction performance.  \citet{Ahrens21} construct a model linking commuting times to rental prices in the Dublin region, with rising rents linked to shorter commuting times showing the importance of spatial proximity in valuation models. \citet{Lyons2019} explores the relationship between the advertised asking price for dwellings that are listed on estate agent platforms and final sale prices, showing they act as a good proxy. However, no property level data in terms of descriptive features are used. \citet{Oust:2020aa} apply a number of spatial modelling techniques to a Norwegian property data set. They use regression kriging and GWR among other models for modelling the spatial component. The best performing model they fit was a GWR model that included measures of repeat sales. By including the repeat sales, they increase their median absolute percentage error from 6.65\% to 6.20\%. 

In the remainder of the section, we present our spatial modelling approach. We harness a Generalised Additive Model (GAM) approach, due to the ease of inclusion of smoothly varying non-linear relationships between the response and independent variables. It also allows us to incorporate spatial random effects. Furthermore, GAMs are extremely computationally efficient and have robust and well understood statistical properties.

\paragraph{Generalised Additive Models}

A Generalised Additive Model, or GAM, is a non-parametric extension of generalised linear models (GLM) with the linear predictor involving a sum of smooth function of covariates (\citet{Wood_GAMBook}). The primary benefit of a GAM based approach over linear models is the flexibility of modelling non-linear relationships between predictor variables and the response, in addition to the interpretability of a standard linear model. Smoothing splines are typically used to model non-linear univariate or bivariate relationships between variables and provide a way to smoothly interpolate between points. Splines allow for a certain degree of localised flexibility, in contrast to the use of higher degree polynomials that would lead to undesirable global effects. 

Spatial GAMs are most commonly used in climate applications due to their computational advantages in modelling large spatio-temporal data sets, for example in the modelling of air pollution, (\citet{ramsay2003exploring} and \citet{Wood_Pollution}), and problems associated with fisheries (\citet{WOOD_Fish}). 
\citet{Pace_1998} was one of the first to showcase the benefit of using a GAM approach in real estate research. In this article, Pace gathered 442 observations over a 6 month period for Memphis, and found that the GAM model improved predictions and reduced the associated prediction errors. \citet{green_gam} have more recently applied GAMs to estimate the effect of green spaces on house prices. \citet{shimizu2014nonlinearity} use various modelling techniques including GAMs when modelling the condominium markets of the Tokyo metropolitan area. When modelling over 9,500 observations their GAM outperformed the linear model but was not the best performing model for out-of-sample data. Their best performing model for out-of-sample data was a continuous dummy variable model (DmM) model.  
\citet{von2015alternative} apply GAMs to a data set containing transactions for single-family homes across the city of Aalborg, Denmark over the period 2000 to 2007. When they compare their GAM models to GLMs and fixed effects models, they find that the inclusion of the spatial process is important to the overall explainability of the model.

GAMs, in a Gaussian context, have the form  
\begin{equation} \label{eq:gam}
\begin{aligned}
\mathrm{log}_e(\boldsymbol{V_i})  = & \boldsymbol{\beta_0} + \sum_k X_{k i} \boldsymbol{\beta_{k}} + \sum_l Z_{l i} \boldsymbol{\gamma_{l}} + f_1(Beds_i; k_1) + \\ 
    & f_2(Baths_i; k_1) + f_3(CBD_i;k_2) + \\ 
    & f_3(Size_i; k_2) + f_4(s_{1i}, s_{2i}; k_3) + \epsilon_i \\
\end{aligned}
\end{equation}

where $\mathrm{log}_e(\boldsymbol{V_i})$  is the log price per $m^2$, and $Z$, $\beta$ and $\gamma$ are as in Equation~\ref{eq1}. 
For `Type', `BER' and `Postcode' we impose a sum to zero constraint with coefficients within each variable relative to the grand mean rather than a base line category, increasing the interpretability of our outputs. 

As opposed to being modelled linearly via $X$ previously, CBD distance and property size are modelled via smoothing splines, $f_{j}$ (\citet{Wood_GAMBook}). For the 1-D splines, we use cubic regression splines with 5 knots for beds and baths, and 20 knots for size and IFSC distance, equally spaced across covariate quantiles.  They are penalized for smoothness using conventional integrated square second derivative cubic spline penalties (\citet{Wood_GAMBook}).

We convert the longitude and latitude recordings to Cartesian coordinates and specify a Gaussian process for the spatial relationship. This requires the specification of a smoothing kernel for the covariance function, based on Euclidean distance between observations. We specify one proposed by \citet{corr_fun}, 
\begin{equation} \label{eq:corr_function}
    C_{\theta}(r) = \sigma_{x}^{2}\left( 1 + \frac{|r|}{\rho}\right)exp\left(-\frac{|r|}{\rho}\right) 
\end{equation}
as this requires only one parameter to be estimated in comparison to spherical and Matérn function approaches. We observed negligible differences in the estimated spatial fields when using this simplified covariance function as compared to more complicated members of the Mat\'{e}rn covariance class. 

Models are fit using the \textit{mgcv} package within the R programming language. A penalized maximum likelihood approach is used in fitting models with smoothness penalties on each of the spline parameters. Models were fit with several different choices of the number of knots for each smoothing spline to determine sensitivity of results to this parameter, which lead to 100 knots for the geo-additive term and is discussed further in Section~\ref{sec:knots}.
We fit four different GAM models with the differences between them outlined in Table \ref{tab:GAMS}, with more model detail provided in Table \ref{tab:Models}. 

\begin{table}[h!]
    \centering
    \caption{GAM Models}
    \label{tab:GAMS}
    \begin{tabular}{c l}
    \hline
    \textbf{GAM} & \textbf{Includes} \\
    \hline
    1 & Linear and smooths, distance variables, no spatial component \\
    2 & Linear for all variables, distance variables, spatial component \\
    3 & Linear and smooths, no distance variables, spatial component \\
    4 & Linear and smooths, distance variables, spatial component\\
    \hline
\end{tabular}
\end{table}

%%%%%%%%%%%%%%%%%%%%%%%%

\section{Results} \label{sec:Results}
For our preliminary models, we used the postcode provided with each property listing. We investigate the impact of address misspecification and resulting bias in Section~\ref{sec:correctpcodes}. For variable selection we follow the approach of \citet[Section~4.6]{GelmanBook}: if a predictor has the expected sign it is retained in the model whether statistically significant or not, otherwise further analysis is carried out on the variable itself to determine its exclusion. We use 5-fold cross validation for model comparison. The accuracy metrics for model comparison are: the prediction intervals; median error percentage; $R^2$ and root mean squared error (RMSE).

\subsection{Selecting the optimal number of knots for the spatial surface}\label{sec:knots}

\begin{figure}[h!]
    \centering
    \includegraphics[width= \textwidth, height = 5cm]{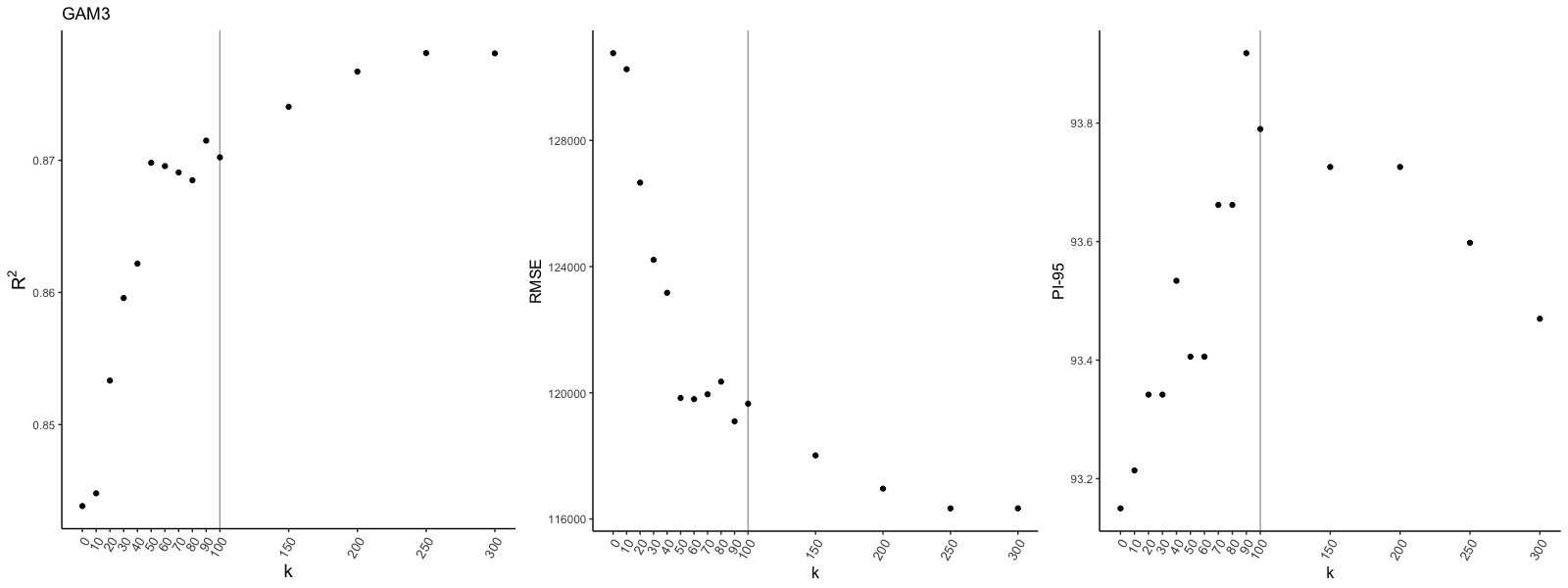}
    \caption{Estimates of $R^2$, RMSE and prediction interval coverage for increasing number of spatial knots.}
    \label{fig:kvals}
\end{figure}

We determine the number of knots required by running the models for increasing knot values and identifying an `elbow' in the plot of $k$ versus accuracy measures. Figure \ref{fig:kvals} shows these plots for GAM 3. From left to right the columns show: $R^2$ against $k$; RMSE against $k$ and 95\% predictive interval coverage against $k$. The prediction intervals measure the uncertainty associated with a price prediction, and enable evaluation of statistical distribution assumptions. Ideally $k$ is as low as possible for computational reasons, but with a requirement to retain satisfactory accuracy statistics - the higher the value of $k$, the longer the models take to fit. From these plots, we highlight the selected value of $k$ = 100 for our spatial surface.

\subsection{Fitted Models and Interpretation}

Table~\ref{tab:accstats_given} presents accuracy estimates for the county of Dublin. These accuracy estimates include median absolute percentage error, percentage of predictions within 10\% of sale price and a number of prediction intervals. We notice higher percentage coverage than expected in the prediction intervals for the linear and GAM 2 models. One reason for this could be the models are not describing the data well, which results with an increase in the error variance parameter and thus the prediction intervals are compensating. As house prices are spatially auto-correlated (\citet{Basu_1998}), residuals of models that do not include a spatial element may retain this. Moran's I is a measure of spatial autocorrelation, and is shown for every model in Table \ref{tab:accstats_given}. Moran's I ranges in values from -1 to 1, with 0 indicating no autocorrelation. Table~\ref{tab:accstats_given} illustrates Moran's I approaching 0 with increasing model complexity.

\begin{table}[h!]
    \centering
    \caption{Accuracy Statistics from 5 Fold Cross Validation}
        \label{tab:accstats_given}
    \resizebox{\textwidth}{!}{\begin{tabular}{r c c c c c c c c }
        \hline
        \textbf{Model} & $\mathbf{R^2}$ & \textbf{RMSE} & \textbf{Median Absolute} & \textbf{Within 10\%} & \textbf{Within 20\%} & \textbf{In 50\% Predictive} &\textbf{ In 95\% Predictive} & \textbf{Moran's} \\
         & & & \textbf{\% Error}& \textbf{of Price} & \textbf{of Price} & \textbf{Interval} & \textbf{Interval} & \textbf{I}\\
        \hline
        Basic Linear Model & 0.73 & €182,508 & 18.32\% & 28.25\% & 54.17\% & 49.7\% & 94.9\% & 0.117  \\
        Linear Model & 0.81 & €154,160 & 12.10\% & 42.99\% & 72.08\% & 54.2\% & 94.6\% & 0.029 \\
        GAM 1 & 0.85 & €132,758 & 11.22\% & 45.81\% & 75.52\% & 47.0\% & 93.1\% & 0.025 \\
        GAM 2 & 0.83 & €146,421 & 9.67\% & 51.75\% & 81.59\% & 52.2\% & 94.8\% & 0.010 \\
        GAM 3 & 0.87 & €123,448 & 8.57\% & 56.37\% & 84.95\% & 48.3\% & 93.9\% & 0.009 \\
        GAM 4 & 0.87 & €123,639 & 8.50\% & 56.37\% & 84.98\% & 48.5\% & 93.9\% & 0.008 \\
        \hline
    \end{tabular}}
\end{table}

Our optimal model is GAM3 which includes a spatial surface, smooths applied to `Size', `No. Beds' and `No. Baths' along with binary variables and no distance metrics, i.e. distance to the Airport and CBD are excluded. We select GAM 3 to be our best model as excluding the distance metrics enhances the interpretability of the spatial surface. GAM 3 has the lowest RMSE shown in Table~\ref{tab:accstats_given}, and has the same values as the GAM 4 model in other measures. 

For GAM 3 we examine the accuracy for closed interval subsets of the data (e.g. between €450,001 and €500,000) and the cumulative median absolute percentage error in Table~\ref{tab:Acc_Intervals}. We note that our accuracy measures remain relatively constant up to the interval €650,001 to €700,000, which when examined cumulatively, accounts for 81.9\% of our data. Up to this interval, the cumulative median absolute percentage error is 8.03\%, 0.54\% lower than the overall median absolute percentage error. Within some intervals, the median absolute percentage error is below 8\% (7.25\% between €300,001 and €350,000; 7.95\% between €350,001 and €400,000; 7.68\% between €500,001 and €550,000; 7.31\% between €600,001 and €650,000). Within Table~\ref{tab:Acc_Intervals} the number of properties within each interval is shown, indicating that larger numbers of properties does not necessarily translate into improved predictions.

\begin{figure}[h!]
    \centering
    \begin{subfigure}{0.475\textwidth}
        \includegraphics[width = \textwidth]{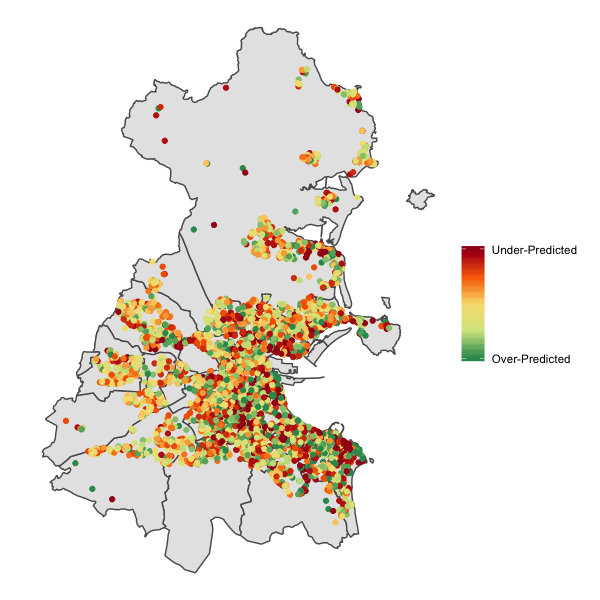}
        \caption{Residuals from 5-fold Cross Validation}
        \label{fig:ResMap}
    \end{subfigure}
    \begin{subfigure}{0.475\textwidth}
        \includegraphics[width = \textwidth]{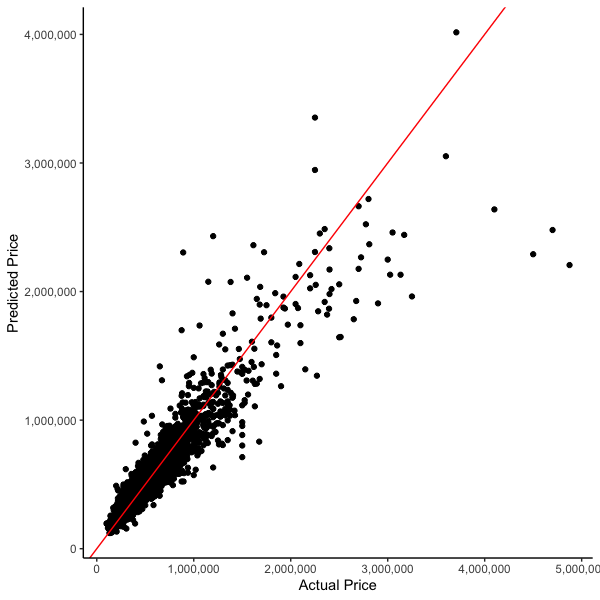}
        \caption{Predicted price against Actual price. Increase in variability with increasing price}
        \label{fig:PredVAct}
    \end{subfigure}
    \caption{Visualising results of GAM 3 model}
    \label{fig:mix1}
\end{figure}

\begin{figure}[h!]
    \centering
    \begin{subfigure}[b]{0.475\textwidth}  
        \centering 
        \includegraphics[width=\textwidth]{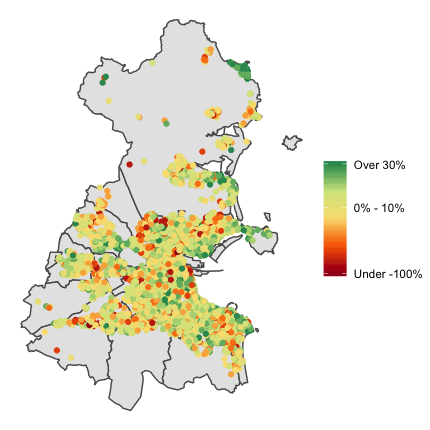}
        \caption{Linear model} \label{fig:errs_lin_map}
    \end{subfigure}
    \hfill
    \begin{subfigure}[b]{0.475\textwidth}
        \centering
        \includegraphics[width=\textwidth]{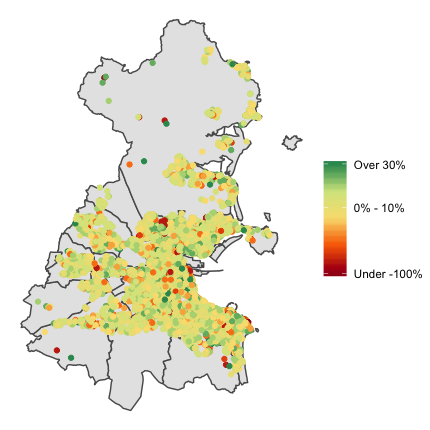}
        \caption{GAM 3 model } \label{fig:errs_gam3_map}
    \end{subfigure}
    \hfill
     \caption{Percentage Errors}
    \label{fig:Percentage_Errs_Map}
\end{figure}

We note the reducing spatial autocorrelation in the residuals of the models that include a spatial surface compared to those that do not. Figure~\ref{fig:ResMap} shows the residuals spatially plotted for the GAM 3 model, which echos the Moran's I statistics that there is little spatial autocorrelation remaining. The percentage errors from the linear model (Figure~\ref{fig:errs_lin_map}) retains a certain degree of spatial structure. We see this with the clustering of similar colours (e.g. dark green in the north east coast), which is less evident for the GAM 3 model (Figure~\ref{fig:errs_gam3_map}).

Although our models give good accuracy measures, there is evidence of heteroscedasticity shown in Figure~\ref{fig:PredVAct}. This is clearly evident in properties that have sold for over €1,000,000. Approximately 5\% of our data set have sale prices over €1,000,000 and this reduces the over all accuracy measures of our models. Figure~\ref{fig:PredVAct} echos our findings from Table~\ref{tab:Acc_Intervals} in which the median absolute percentage error increases substantially for properties with a sales price of €1,000,000 or more, decreasing our predictive performance and increasing the variability in these predictions. 

\begin{figure}[h!]
    \centering
    \includegraphics[width = \textwidth, height = 13cm]{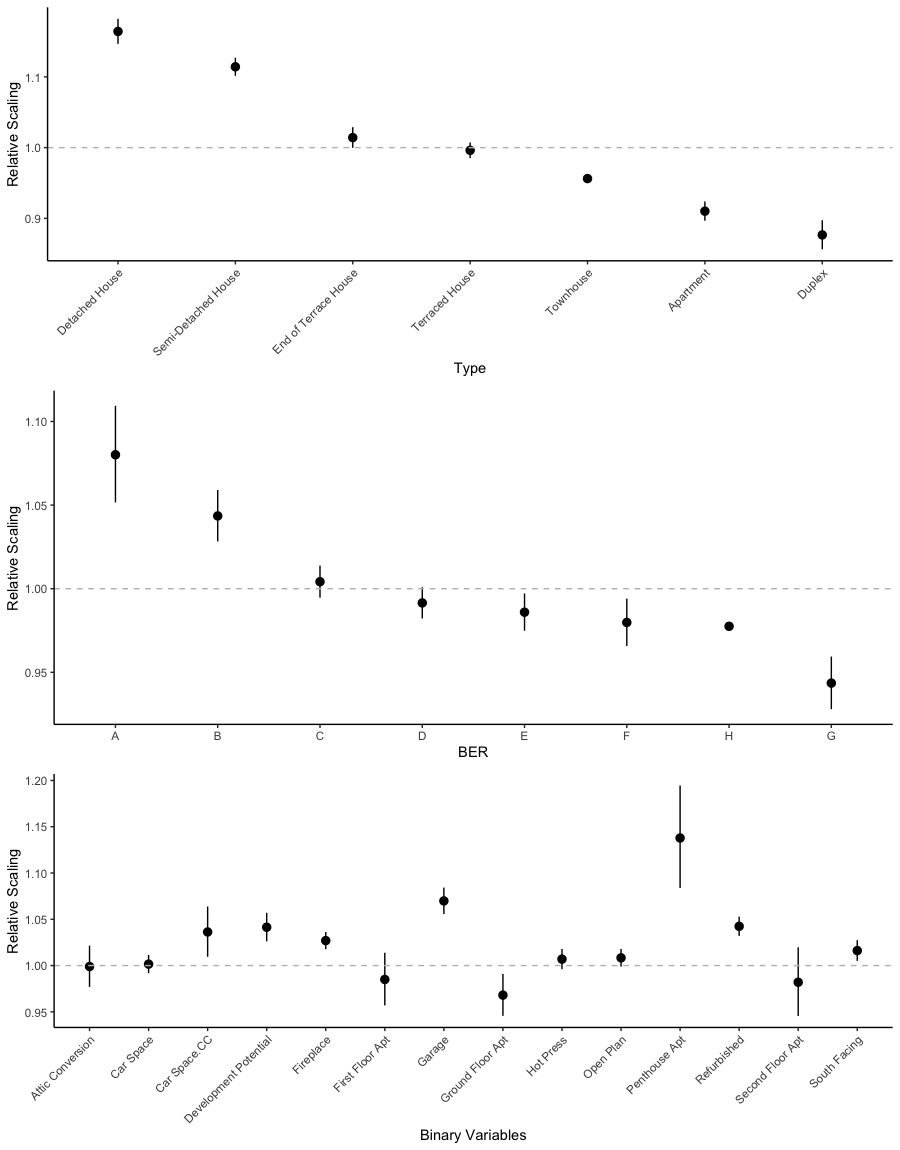}
    \caption{Model coefficients for GAM 3 model, linear variables divided into three groups: `Type', `BER' and `Binary Variables'}
    \label{fig:mod_coeffs}
\end{figure}

\paragraph{Interpretation of Coefficients}
In the following we focus on the outputs of the GAM 3 model. The model parameter coefficients are relative scalings. For categorical variables, rather than being compared to a base category we constrain with respect to to the overall grand mean across levels within the variable. The coefficients can be observed in Figure~\ref{fig:mod_coeffs}, where the linear coefficients are broken into three plots, showcasing the coefficients for `Type', `BER' and `Binary Variables' separately. Table~\ref{tab:mod_coeffs} gives the estimated coefficients and their associated 95\% confidence intervals. In Figure~\ref{fig:mod_coeffs} and Table~\ref{tab:mod_coeffs}, a relative scaling of 1 indicates no impact from that variable, a relative scaling less than 1  shows a reduction in value and greater than 1 an increase in value. To aid visualisation in Figure~\ref{fig:mod_coeffs}, a relative scaling of 1 is highlighted by a grey dashed line.  

These plots enable us to create a hierarchy of property types and BERs. This aids in quantifying the additional premiums associated between different categories of these variables. This enables us to assess the premiums for various combinations, not just relative to a base line category. For example, we estimate that a detached house has a coefficient value between 1.15 and 1.18, with mean 1.16, and a semi-detached house has a coefficient value between 1.10 and 1.13 with mean 1.11. Consequently this would indicate that the premium for a detached house over a semi-detached equivalent would be 4\% $(1-\frac{1.11}{1.16})$. Since our coefficients are relative scalings, we can also calculate the premium for an apartment over a duplex equivalent to be 3\% $(1-\frac{0.88}{0.91})$. Similar calculations for comparisons can be made for BER ratings - upgrading a property from a G rating to an A rating increases the value by 13\% $(1-\frac{0.94}{1.08})$. We also estimate the premium for a renovated property to be between 3\% and 5\% over a non-renovated equivalent. In our model we include both the `Car Space' and `Car Space.CC' terms, and although the coefficient of `Car Space' is 1, implying no impact, `Car Space.CC' has a coefficient of 1.04 signifying an additional 4\% premium for properties with a car parking within the city centre. Unsurprisingly, having a garage adds a significant premium of between 6\% and 8\%. However, the largest estimated premium is associated with `Penthouse Apartment' which takes a value between 8\% and 19\%. 
From these estimates, we can indicate to homeowners the possible additional value undertaking home improvements will add to their property. These estimates also lend themselves to policy makers when allocating funding for people to improve their homes energy performance or BER. 

\begin{figure}[h!]
    \centering
    \begin{subfigure}{0.3\textwidth}
        \includegraphics[width = \textwidth]{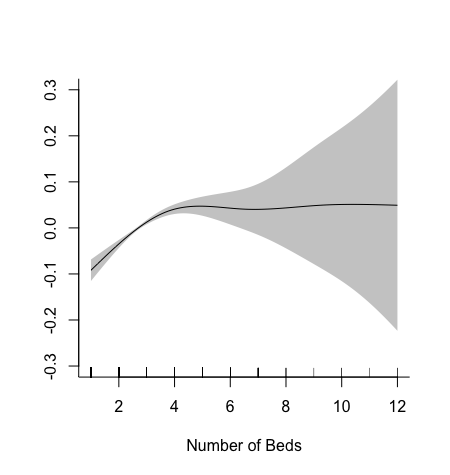}
        \label{fig:BedsSmooth}
    \end{subfigure}
    \begin{subfigure}{0.3\textwidth}
        \includegraphics[width = \textwidth]{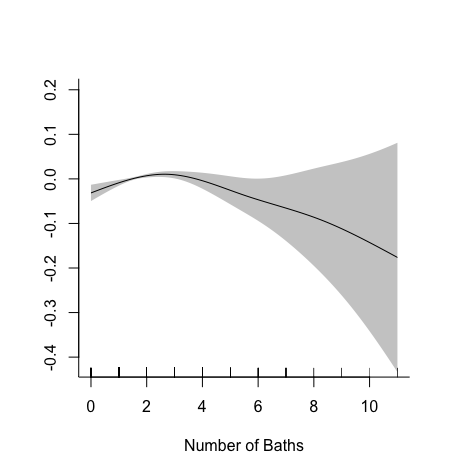}
        \label{fig:BathsSmooth}
    \end{subfigure}
    \begin{subfigure}{0.3\textwidth}
        \includegraphics[width = \textwidth]{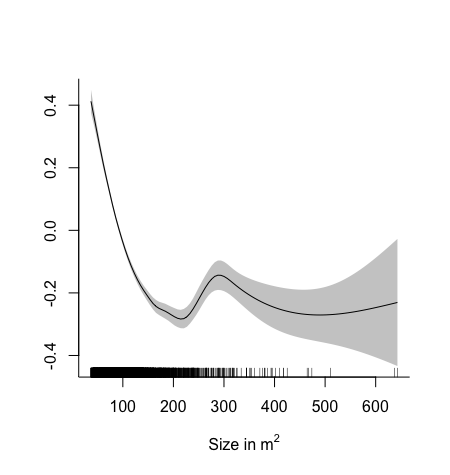}
        \label{fig:SizeSmooth}
    \end{subfigure}
    \caption{Visualising 1D smooths from GAM 3 model}
    \label{fig:smooths3}
\end{figure}

\paragraph{Investigating Smooths}
Although the coefficients of the smooths do not have the same level of interpretability as those of the linear terms, we can visualise the smooths (Figure~\ref{fig:smooths3}) and interpret the relationships.

Starting from the leftmost plot in Figure~\ref{fig:smooths3}, we see that the effect of beds, holding all other variables constant, increases and then plateaus in value add at 4 beds. This would imply that there is little difference in effect of additional beds  after 4. However, the uncertainty intervals are large after 6 beds given a paucity of properties with beds exceeding this number. Similarly, keeping all other variables constant, the relationship between value and baths initially increases until it reaches 2, and then declines. This decline is most likely associated with old properties than have been split into several multi-occupancy units and require large investments to bring them up to modern standards. The most interesting relationship is perhaps between property value and size, when all other covariates remain unchanged. Smaller properties, up to 150m$^2$, are more expensive than larger ones on a per $m^2$ basis. The effect would appear to plateau at 200m$^2$, but for an unexpected increase at 300m$^2$. Investigating the 88 properties that have sizes between 250m$^2$ and 400m$^2$ reveals that the majority are large Georgian or Victorian properties in extremely affluent areas of the city and highly sought after. 

\subsection{Machine Learning Based Estimates} \label{sec:MLResults}
For our small data set, we fit three different machine learning approaches: K-Nearest Neighbours, Decision Trees and Random Forests. Our aim is to evaluate the performance of these approaches on small data sets, as their performance on large data sets has been well documented across the real estate literature.

The nearest neighbours algorithm we implement (Algorithm~\ref{alg::nns}) is on a price per $m^2$ basis, controlling for property size. The results of this approach are shown in Table~\ref{tab:nearest_neighbours_ppm_tab}. 

\begin{table}[h!]
    \centering
    \caption{Accuracy statistics for nearest neighbours approach for price per $m^2$}
    \label{tab:nearest_neighbours_ppm_tab}
    \begin{tabular}{c c c c }
    \hline
    \textbf{Number of } & \textbf{Median Absolute} & \textbf{Within 10 \%} & \textbf{Within 20 \%}\\
    \textbf{Neighbours} & \textbf{\% Error}& \textbf{of Price} & \textbf{of Price}\\
    \hline
    3   & 11.30\%  & 45.37\% & 73.91\% \\
    5   & 10.98\%  & 46.47\% & 75.29\% \\
    7   & 10.95\%  & 46.70\% & 75.40\% \\ 
    9   & 10.92\%  & 46.72\% & 75.29\% \\
%    11  & 10.90\%  & 46.91\% & 75.60\% \\
    \hline
\end{tabular}
\end{table}

The statistics seen in Table~\ref{tab:nearest_neighbours_ppm_tab} give similar or improved statistics compared to the linear models from Table~\ref{tab:accstats_given}, but on a whole under-perform when compared to the more flexible GAM models. There are also indications that increasing the number of neighbours does not necessarily mean improved accuracy.

We trained both decision tree and random forest models as an alternative. Results from the implementation of two decision trees and two random forests are presented in Table~\ref{tab:trees}. `Tree 1' is a decision tree that includes all variables apart from `Longitude' and `Latitude', `Tree 2'  is a decision tree that includes all variables including `Longitude' and `Latitude' and `Forest 1' and `Forest 2' follow the same convention.
 
\begin{table}[h!]
    \caption{Results from decision tree and random forest approaches}
    \label{tab:trees}
    \centering
    \resizebox{\textwidth}{!}{\begin{tabular}{l c c c c c c}
        \hline
        \textbf{Model} & \textbf{R$^2$} &\textbf{RMSE} & \textbf{Median Absolute} & \textbf{Within 5\%} & \textbf{Within 10\%} & \textbf{Within 20\%}\\
         & & & \textbf{\% Error} & \textbf{of Price} & \textbf{of Price} & \textbf{of Price}\\
        \hline
        Tree 1      & 0.80 & €149,444 & 13.48\% & 20.22\% & 39.27\% & 67.49\% \\
        Tree 2      & 0.81 & €148,977 & 13.32\% & 20.56\% & 39.13\% & 68.33\% \\
        Forest 1    & 0.87 & €122,937 & 9.54\%  & 28.01\% & 51.71\% & 80.07\% \\
        Forest 2    & 0.88 & €117,328 & 8.82\%  & 30.34\% & 55.38\% & 83.53\% \\
        \hline
    \end{tabular}}
\end{table}

From Table~\ref{tab:trees}, we see that the random forests are marginally more accurate at predicting the mean (R$^2$ and RMSE) than our GAM models from Table~\ref{tab:accstats_given}, however are less accurate in prediction interval coverage. 
Although we can compare the results of the random forests and GAM models, in taking a GAM approach we reap the benefit of being able to interpret our models and can assess the impacts of specific variables.

\subsection{The Impact of Address Mislabelling?} \label{sec:correctpcodes}

\begin{figure}[h!]
    \begin{floatrow}
    \ffigbox{%
      \includegraphics[width = 0.4\textwidth, height = 0.335\textwidth]{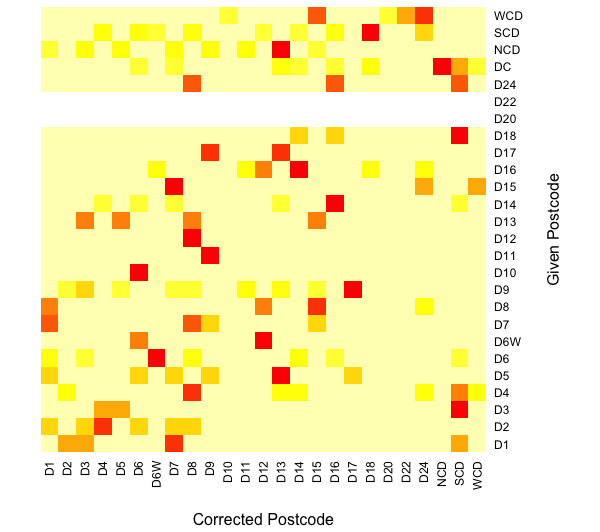}
    }{%
      \caption{Heatmap of the changes that occur in postcodes}%
      \label{fig:changes}%
    }
    \capbtabbox{%
      \resizebox{0.465\textwidth}{!}{\begin{tabular}{c c c }
        \hline
        \textbf{Given Postcode} & \textbf{Corrected Postcode} & \textbf{Occurrences} \\
        \hline
        Dublin 5 & Dublin 13 & 3 \\
        Dublin 6 & Dublin 6W & 12 \\
        Dublin 6W & Dublin 12 & 3 \\
        Dublin 9 & Dublin 3 & 3\\            
        Dublin 9 & Dublin 17 & 10 \\
        Dublin 11 & Dublin 9 & 7 \\
        Dublin 14 & Dublin 16 & 19 \\
        Dublin 16 & Dublin 14 & 5 \\
        North County Dublin & Dublin 15 & 9 \\
        South County Dublin & Dublin 4 & 27 \\
        South County Dublin & Dublin 18 & 101\\
        South County Dublin & Dublin 24 & 32 \\
        West County Dublin & Dublin 15 & 38 \\
        West County Dublin & Dublin 22 & 24 \\
        West County Dublin & Dublin 24 & 45 \\
        \hline
    \end{tabular}}
    }{
      \caption{15 Postcode Changes Investigated}%
      \label{tab:Changes}%
    }
    \end{floatrow}
\end{figure}

The initial analysis used the given postcodes on the property listings. 
However, some properties had given the neighbouring, typically more affluent, postcode rather than their actual one. We assess the impact of postcode misspecification by correcting the postcodes and re-running our analysis. 16\% of properties in the data set (833 properties) gave the incorrect postcode in their advertisement. We investigate 15 postcode changes (outlined in Table~\ref{tab:Changes}), based on a cross tabulation of the postcodes, shown in Figure~\ref{fig:changes}, and prior knowledge of the Dublin property market. In Figure~\ref{fig:changes}, the white lines for D20 and D22 indicate no changes in postcode labeling was required. In the models `GAM 5' and `GAM 6' in Table~\ref{tab:accstats}, the postcode changes are modelled by adding 15 dummy variables to previous models, one for each postcode change. GAM 5 and GAM 6 are the GAM 3 and GAM 4 models respectively, with the postcodes corrected.

\begin{figure}[h!]
    \centering
    \includegraphics[width = \textwidth, height = 9cm]{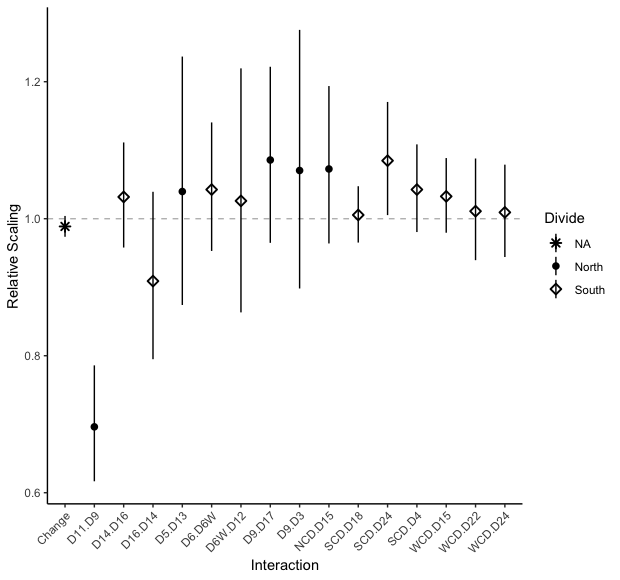}
    \caption{Estimates of Postcode Changes}
    \label{fig:Pcode_Changes}
\end{figure}

Figure~\ref{fig:Pcode_Changes} presents the estimates for the 15 postcode changes. The labels can be read as `Given Postcode.Corrected Postcode', e.g. ``D11.D9" is a property listed as D11 but with correct postcode D9. The majority of the estimated coefficients are positive, indicating a premium in sales price in each case where giving a neighbouring, more affluent postcode. 
It is clear that ``D11.D9" and ``D16.D14" are different to the other changes estimated, with scalings less than 1. These properties have listed postcodes with a lower median price (less affluent) than their correct postcode. For properties incorrectly listing a more affluent postcode, there is evidence of an additional premium being added to property values - for 13/16 postcode changes the mean scaling estimates are greater than 1. The estimates have larger confidence intervals than those seen previously in Figure~\ref{fig:mod_coeffs}, as there are fewer transactions to estimate from. 

\begin{figure}[h!]
    \centering
    \includegraphics[width = \textwidth, height = 9cm]{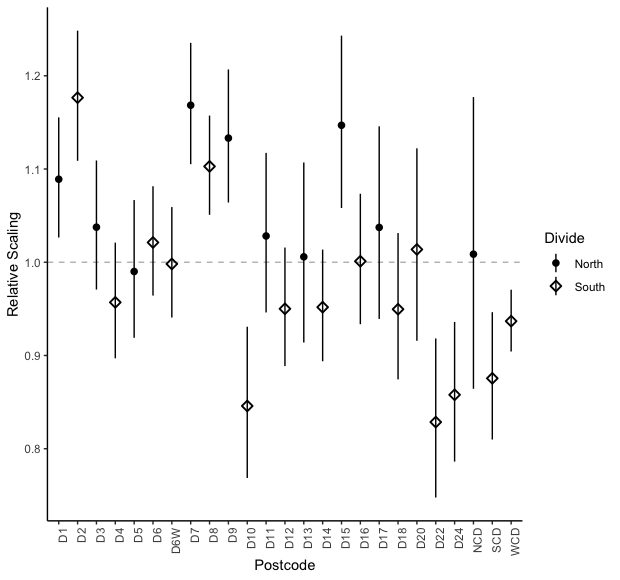}
    \caption{Dot and Whisker Plot for Corrected Postcodes}
    \label{fig:P_DWPlot}
\end{figure}

 In Figure~\ref{fig:P_DWPlot} we present the estimated scalings associated with the corrected postcodes. If the underlying spatial surface captured all spatial structure in the data, the postcodes should be labels alone, with no obvious patterns. However we note there is substantial evidence of a strong bias in postcode effects. The postcodes D7, D8 and D9 have extremely significant impacts on price with these representing popular, up-and-coming areas reflected in scaling estimates much greater than 1. D10, D22 and D24 all have scalings substantially less than 1 which would suggest these areas are undervalued relative to their spatial location. This is perhaps due to negative connotations regarding areas within these postcodes that are socially deprived and underprivileged. Table~\ref{tab:CP_Tab1} lists the mean estimate and confidence intervals for the postcodes.

Overall, there is little difference in accuracy between using given postcodes and corrected postcodes (Table~\ref{tab:accstats_given} and Table~\ref{tab:accstats}). 
We do see improvements in the median absolute percentage error as it reduces for all models and there is a rise in the proportions within 10\% and 20\% of sale price. Though we do see improvements with our predictions when using the corrected postcodes, these improvements are relatively minor.
    
\begin{table}[h!]
    \centering
    \caption{Accuracy Statistics for 5 Fold Cross Validation, Corrected Postcodes}
        \label{tab:accstats}
    \resizebox{\textwidth}{!}{\begin{tabular}{r c c c c c  c c c }
        \hline
        \textbf{Model} & $\mathbf{R^2}$ & \textbf{RMSE} & \textbf{Median Absolute} & \textbf{Within 10 \%} & \textbf{Within 20 \%} & \textbf{In 50\% Predictive} &\textbf{ In 95\% Predictive} & \textbf{Moran's} \\
         & & & \textbf{\% Error}& \textbf{of Price} & \textbf{of Price} & \textbf{Interval} & \textbf{Interval} & \textbf{I}\\
        \hline
        Basic Linear Model & 0.74 & €182,526 & 18.35\% & 28.34\% & 54.24\% & 49.8\% & 95.0\% & 0.117\\
        Linear Model & 0.81 & €150,767 & 11.65\% & 43.84\% & 73.14\% & 54.9\% & 94.8\% & 0.025\\
        GAM 1 & 0.85 & €131,775 & 10.93\% & 46.72\% & 76.75\% & 46.8\% & 93.5\% & 0.025\\
        GAM 2 & 0.83 & €146,719 & 9.62\% & 51.44\% & 81.99\% & 52.3\% & 94.8\% & 0.010\\
        GAM 3 & 0.87 & €123,658 & 8.45\% & 57.16\% & 85.10\% & 48.5\% & 94.1\% & 0.009\\
        GAM 4 & 0.87 & €124,640 & 8.54\% & 56.57\% & 85.22\% & 48.6\% & 94.1\% & 0.009\\
        GAM 5 & 0.87 & €123,713 & 8.48\% & 57.22\% & 85.02\% & 48.6\% & 94.1\% & 0.009 \\
        GAM 6 & 0.87 & €124,717 & 8.63\% & 56.43\% & 84.97\% & 48.7\% & 94.1\% & 0.008 \\
        \hline
    \end{tabular}}
\end{table}

\begin{figure}[h!]
    \centering
    \includegraphics[width = \textwidth]{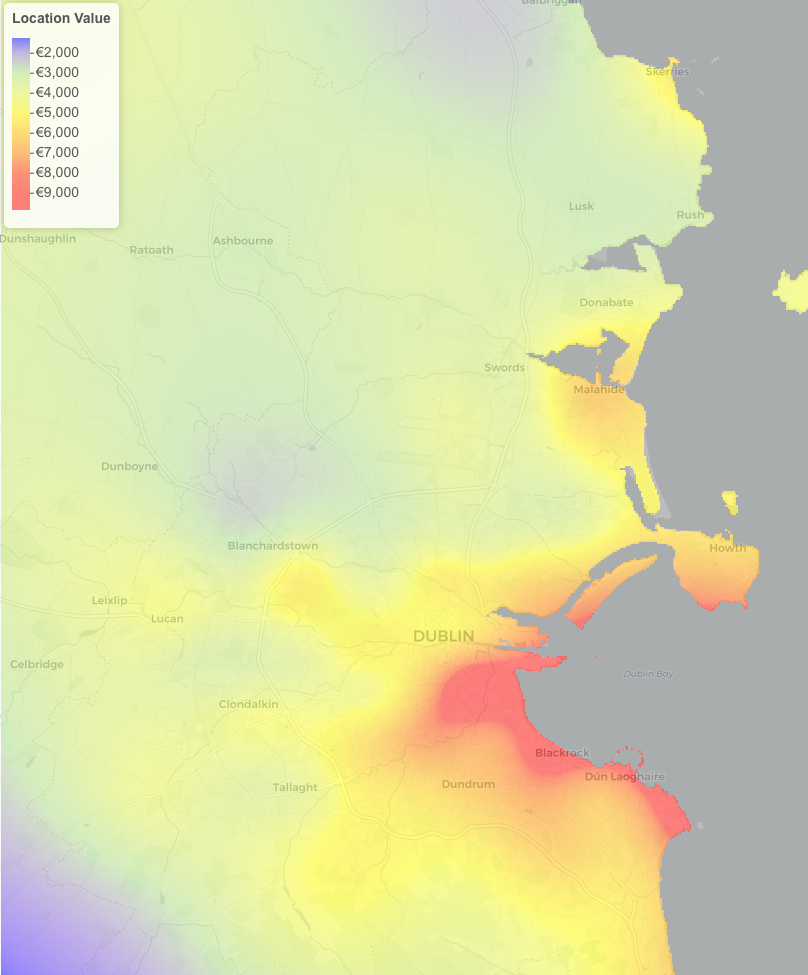}
    \caption{Spatial surface on map of Dublin}
    \label{fig:heatmap}
\end{figure}

We can visualise and explore the Gaussian Process smooth used to model the spatial surface. We overlay this surface on a map of Dublin and investigate whether the surface makes sense in the context of the Dublin residential property market. Figure \ref{fig:heatmap} shows the spatial surface which gives the Location Values for the GAM 5 model. The colour scale in Figure~\ref{fig:heatmap} uses blue for low values and red for high values. 
As anticipated, the higher Location Values are in the more affluent areas along the south coast line. There are some other localised `hot-spots', which are other areas that are affluent or in high demand. One such localised `hot-spot' is in the Castleknock region, just below ``Blanchardstown" in Figure~\ref{fig:heatmap}. This would indicate that this area is more valuable than its surroundings.  
These Location Values are the associated price per square metre premiums associated to specific GPS coordinates. 

\section{A Proxy Site Valuation Model}

In an Irish context, \citet{Collins2011}  investigate the design and implementations of an SVT for Ireland, noting the main impediment is the difficulty of providing land valuations.  \citet{OHanlon2011} develop a rolling hedonic regression model based on stamp duty returns to the Irish Revenue Commissioners. They note the substantial issue of data quality with large amounts of missing information. \citet{Maguire2016} also provide an overview of a number of different approaches to the construction of a price register in an Irish context. There is no published valuation models in an Irish context that utilise property specific information in their pricing models. 

One potential use of Figure~\ref{fig:heatmap} is in assisting in a site valuation tax calculation. To do this, the Location Values need to be scaled by the minimum Location Value. This would result in a Scaling Value per GPS coordinate, ranging in value from 1 to 8. These scaling values could be used to multiply a baseline tax per acre provided by the Irish government. The requirements of this approach would then be limited to the scaling values and the site size of a property.

\begin{equation} \label{eq:propsitetax}
    \text{Site Value Tax} = \text{Scaling Value} \times \text{Site Size}  \times \text{Baseline} %s(\mathrm{Lat, Lon})
\end{equation}
Alternatively, bands based on the Scaling Values could be used rather than the individual Scaling Values.

As mentioned by \citet{Collins2011}, unlike other property types both apartments and duplexes share a site. Our first proposal is similar to that proposed by \citet{Collins2011}: the site is taxed separately to that of the individual apartments. In essence, the site is taxed as in Equation~\ref{eq:propsitetax} and the apartments are taxed using the current local property tax calculations. 

duplexes, allowances can be made. 
The second is to scale the previous equation by the number of apartments on the site, thus giving a `site value' for each apartment. So rather than the site tax falling to the owner(s) of the complex, it would be included in the tax owed by each individuals apartment.
\begin{equation}\label{eq:sitetaxapts}
    \text{Site Value Tax} = \frac{\text{Scaling Value} \times \text{Site Size}  \times \text{Baseline}}{\text{Number of Apartments}}
\end{equation}

In Ireland, all residential properties owe what is known as the local property tax, LPT. It is calculated as a percentage of the resale value of the house. The site value tax calculations shown in Equations~\ref{eq:propsitetax} and \ref{eq:sitetaxapts} could be included in this tax to potentially make it more equitable.

\section{Discussion and Conclusion} \label{sec:conclusions}
This article considers two different modelling techniques for property price prediction: hedonic regression and a flexible approach using GAMs. We account for non linear relationships between the outcome (price) and a number of different housing attributes. This also enables the inclusion of a Gaussian Process smooth for estimating the Location Values. Our results show a reduction in median absolute percentage error with increasing model complexity: 12.10\% for a hedonic model; 9.67\% for a linear model with a spatial surface and 8.57\% by modelling some features with cubic splines. We also examine the accuracy within closed intervals, which illustrated a median absolute percentage error of 8.03\% for 81.9\% of the data. 

One unique feature of our data set is the inclusion of a description variable. This variable contains valuable information including descriptive features of the property and its surroundings. We extract different property features from the textual description, using a simple case insensitive text mining technique, which we include in our models. This allows us to account for features from existing literature and those specific to the Dublin housing market. Including socio-demographic data from the Economic and Social Research Institute (ESRI) may improve the predictive accuracy of our models, but weaken the interpretability of the models and coefficients. Though there may be a correlation between peoples wealth status and the price of the properties in the area where they live, this might not be the causal link.  
As our models separate the impacts of housing features and spatial location, we evaluate estimates of these features that should not contain any residual effect from location. This potentially has widespread applicability, as we imagine that the estimates of the property features would remain constant, thus only requiring an updated Location Value surface. This surface would require a smaller subset of transactions that refitting the model itself. 

Machine learning approaches, such as random forests and decision trees, are being used more frequently for price prediction. Though they are relatively easy to fit, they do not provide probability-based uncertainty intervals and do not have the interpretability of the models presented in this article.
Random forest and decision trees may not extrapolate well to areas outside of the city where housing turnover is low and where the advantages of a statistical modelling based framework will become more pronounced.

If our models were to be expanded to the Republic of Ireland, we would have no choice but to use the postcode that was given in the listing. Apart from Cork City and County Dublin, the rest of the island has no well defined postcodes. This disables us from correcting the postcodes of the majority of the island, thus being unable to correct for address misspecification. 
With models expanded to the whole Republic of Ireland, they could potentially be used to improve property tax calculations and site value estimations.
Current tax calculations are based on the homeowners valuation of the property. Since the value of the tax is based on the value of the property, there is no incentive to upgrade or update the property. By using parts of our models, taxes would not solely be based on property value, but include some element of spatial location scaling. 

\begin{acknowledgements}
The authors gratefully acknowledge the helpful comments from the reviewers, which have greatly improved this work.

\noindent    This publication has emanated from research conducted with the financial support of Science Foundation Ireland under Grant Number 18/CRT/6049. For the purpose of Open Access, the author has applied a CC BY public copyright license to any Author Accepted Manuscript version arising from this submission.
\end{acknowledgements}



\printbibliography
   % name your BibTeX data base

\pagebreak
\section{Appendix}

\begin{table}[h!]
    \centering
    \caption{Variables used in each model.  `L' represents linear variables, `S' those modeled with splines, `GP' indicates a Gaussian Process smooth and $^*$ indicates variables that are statistically significant at 5\% or better in the model.}
    \label{tab:Models}
    \begin{adjustbox}{width=1\textwidth}
    \begin{tabular}{l l l l l l | l l l l l l}
    \hline
     & & & \textbf{\underline{Given Postcode}} & & & & & \textbf{\underline{Corrected Postcode}} & & &\\
     & & & & & & & & & & & \\
    \textbf{Variable} & \textbf{Linear} &  \textbf{GAM 1} & \textbf{GAM 2} & \textbf{GAM 3} & \textbf{GAM 4} &  \textbf{GAM 1} & \textbf{GAM 2} & \textbf{GAM 3} & \textbf{GAM 4} & \textbf{GAM 5} & \textbf{GAM 6}\\
    \hline
    Attic Conversion	    &	L$^*$&	L&	L$^*$&	L&	L   &L   &L$^*$   &L   &L	&L   &L	\\
    South Facing	        &	L$^*$&	L$^*$&	L$^*$&	L$^*$&	L$^*$   &L$^*$   &L$^*$   &L$^*$   &L$^*$	&L$^*$   &L$^*$	\\
    Development Potential	&	L$^*$&	L$^*$&	L$^*$&	L$^*$&	L$^*$   &L$^*$   &L$^*$   &L$^*$   &L$^*$	&L$^*$   &L$^*$	\\
    Fireplace	            &	L$^*$&	L$^*$&	L$^*$&	L$^*$&	L$^*$   &L$^*$   &L$^*$   &L$^*$   &L$^*$	&L$^*$   &L$^*$	\\
    Central Heating	        &	&	&	&	&	    &    &	  &    &    &    &  \\
    Hot Press	            &	L&	L&	L&	L&	L   &L   &L   &L   &L	&L   &L	\\
    Immersion	            &	L&	L&	&	&	    &	 &    &    &    &    &	\\
    Garage	                &	L$^*$&	L$^*$&	L$^*$&	L$^*$&	L$^*$   &L$^*$   &L$^*$   &L$^*$   &L$^*$	&L$^*$   &L$^*$	\\
    Garden	                &	&	&	&	&	    &    &	  &    &    &    &  \\
    Refurbished             &	L$^*$&	L$^*$&	L$^*$&	L$^*$&	L$^*$   &L$^*$   &L$^*$   &L$^*$   &L$^*$	&L$^*$   &L$^*$	\\
    Car Space	            &	L&	&	&	L&	    &    &L   &L   &L	&L   &L	\\
    Cul de Sac 	            &	&	&	&	&	    &	 &    &    &    &    &  \\
    Open Plan	            &	L&	L&	L&	L&	L   &L$^*$   &L   &L   &L	&L   &L	\\
    Ground Floor Apartment	&	L&	L&	L&	L$^*$&	L$^*$   &L$^*$   &L   &L$^*$   &L$^*$	&L$^*$   &L$^*$	\\
    1st Floor Apartment 	&	L&	L&	L&	L&	L   &L   &L   &L   &L	&L   &L	\\
    2nd Floor Apartment 	&	L&	L&	L&	L&	L   &L   &L   &L   &L	&L   &L	\\
    Penthouse Apartment 	&	L&	L$^*$&	L$^*$&	L$^*$&	L$^*$   &L$^*$   &L   &L$^*$   &L$^*$	&L$^*$   &L$^*$	\\
    Type	                &	L$^*$&	L$^*$&	L$^*$&	L$^*$&	L$^*$  &L$^*$   &L$^*$   &L$^*$   &L$^*$	&L$^*$   &L$^*$	\\
    BER 	                &	L$^*$&	L$^*$&	L$^*$&	L$^*$&	L$^*$   &L$^*$	 &L$^*$   &L$^*$   &L$^*$   &L$^*$   &L$^*$	\\
    Given Postcode          &	L$^*$&	L$^*$&	L$^*$&	L$^*$&	L$^*$   &    &    &    &    &    &	\\
    Corrected Postcode      &   &   &   &   &       &L$^*$   &L$^*$   &L$^*$   &L$^*$   &L$^*$   &L$^*$	\\
    Postcode Change         &   &   &   &   &       &    &    &    &    &L$^*$   &L$^*$	\\
    Park Distance	        &	L&	L&	&	&	 	&L   &    &    &L$^*$   &    &L \\
    IFSC Distance 	        &	L$^*$&	S&	L&	&	S	&S   &L   &    &S   &    &S \\
    Airport Distance 	    &	L$^*$&	L$^*$&	L&	&	L	&L$^*$   &L   &    &L   &    &L \\
    City Centre Distance	&	&	&	&	&		&    &    &    &    &    &  \\
    DART Distance 	        &	L$^*$&	L$^*$&	L&	&	L	&L$^*$   &L   &    &L   &    &L \\
    LUAS Distance	        &	L&	L$^*$&	L&	&	L$^*$	&L   &L   &    &L$^*$   &    &L$^*$ \\
    Size	                &	L$^*$&	S&	L$^*$&	S&	S	&S   &L$^*$   &S   &S   &S   &S \\
    Garden.CC	            &	L&	&	L&	&		&    &L   &    &    &    &  \\
    CarSpace.CC 	        &	L&	L$^*$&	L$^*$&	L$^*$&	L   &L   &L$^*$   &L$^*$   &L$^*$	&L$^*$   &L$^*$	\\
    Beds	                &	&	S&	L$^*$&	S&	S	&S   &L$^*$   &S   &L$^*$   &S   &L$^*$ \\
    Baths	                &	&	L&	L&	S&	S	&S   &L   &S   &L   &S   &L  \\
    Spatial Surface 	    &	&	&	GP&	GP&	GP 	&    &GP  &GP  &GP  &GP  &GP\\
    \hline
    \end{tabular}
    \end{adjustbox}
\end{table}

\begin{table}[h!]
    \caption{A set of accuracy statistics for intervals for GAM 3 model}
    \label{tab:Acc_Intervals}
    \centering
    \resizebox{\textwidth}{!}{\begin{tabular}{l c c c c c c }
        \hline
        & \textbf{Number of } & \textbf{Median Absolute} & \textbf{Within 5\%} & \textbf{Within 10\%} & \textbf{Within 20\%} & \textbf{Cumulative Median}\\ 
        & \textbf{Properties} & \textbf{\% Error} & \textbf{of Price} & \textbf{of Price} & \textbf{of Price} & \textbf{Absolute \% Error}\\
        \hline
        Under €250,000          & 703   & 9.23\%    & 28.02\% & 52.92\% & 79.94\% & 9.23\% \\ 
        €250,001 - €300,000     & 707   & 8.08\%    & 32.53\% & 58.98\% & 86.42\% & 8.75\% \\ 
        €300,001 - €350,000     & 719   & 7.25\%    & 35.33\% & 61.06\% & 86.93\% & 8.20\% \\ 
        €350,001 - €400,000     & 582   & 7.95\%    & 34.88\% & 60.48\% & 89.18\% & 8.07\% \\
        €400,001 - €450,000     & 479   & 8.14\%    & 30.48\% & 60.54\% & 89.56\% & 8.10\% \\
        €450,001 - €500,000     & 377   & 8.15\%    & 32.89\% & 57.82\% & 87.53\% & 8.10\% \\ 
        €500,001 - €550,000     & 272   & 7.68\%    & 35.66\% & 61.76\% & 88.97\% & 8.06\% \\ 
        €550,001 - €600,000     & 236   & 8.52\%    & 29.66\% & 57.20\% & 88.56\% & 8.07\% \\ 
        €600,001 - €650,000     & 192   & 7.31\%    & 36.98\% & 61.98\% & 86.46\% & 8.03\% \\ 
        €650,001 - €700,000     & 190   & 9.34\%    & 29.47\% & 54.21\% & 87.37\% & 8.10\% \\ 
        €700,001 - €800,000     & 235   & 10.22\%   & 25.11\% & 48.94\% & 84.26\% & 8.20\% \\ 
        €800,001 - €900,000     & 158   & 10.90\%   & 22.78\% & 45.57\% & 76.58\% & 8.27\% \\ 
        €900,001 - €1,000,000   & 92    & 10.30\%   & 23.91\% & 50.00\% & 78.26\% & 8.31\% \\ 
        €1,000,001 - €1,100,000 & 61    & 12.50\%   & 24.59\% & 40.98\% & 78.69\% & 8.33\% \\ 
        €1,100,001 - €1,500,000 & 121   & 15.80\%   & 18.18\% & 33.88\% & 65.29\% & 8.46\% \\ 
        €1,500,001 - €2,000,000 & 36    & 15.43\%   & 16.67\% & 33.33\% & 61.11\% & 8.52\% \\ 
        €2,000,001 - €3,500,000 & 42    & 18.65\%   & 19.05\% & 28.57\% & 57.14\% & 8.56\% \\ 
        €3,500,001 - €5,000,000 & 6     & 44.71\%   & 0.00\%  & 0.00\%  & 16.67\% & 8.57\% \\ 
   \hline
\end{tabular}}
\end{table}

\begin{table}[h!]
    \centering
    \begin{tabular}{l c c}
        \hline
        \textbf{Variable} & \textbf{Estimate} & \textbf{95\% Confidence Interval}  \\
        \hline
        \textit{Type} & & \\
        Apartment               & 0.91 & [0.90, 0.92] \\
        Detached House          & 1.16 & [1.15, 1.18] \\
        Duplex                  & 0.88 & [0.86, 0.90] \\
        End of Terrace House    & 1.01 & [1.00, 1.03] \\ 
        Semi-Detached House     & 1.11 & [1.10, 1.13] \\
        Terraced House          & 1.00 & [0.99, 1.01] \\
        Townhouse               & 0.96 & [0.96, 0.96] \\
        \textit{BER}            & & \\
        A                       & 1.08 & [1.05, 1.11] \\
        B                       & 1.04 & [1.03, 1.06] \\
        C                       & 1.00 & [0.99, 1.01] \\
        D                       & 0.99 & [0.98, 1.00] \\
        E                       & 0.99 & [0.97, 1.00] \\
        F                       & 0.98 & [0.97, 0.99] \\
        G                       & 0.94 & [0.93, 0.96] \\
        Exempt                  & 0.98 & [0.98, 0.98] \\
        \textit{Binary}          & & \\
        Attic Conversion        & 1.00 & [0.98, 1.02] \\
        Car Space               & 1.00 & [0.99, 1.01] \\
        Car Space.CC            & 1.04 & [1.01, 1.06] \\
        Development Potential   & 1.04 & [1.03, 1.06] \\
        Fireplace               & 1.03 & [1.02, 1.04] \\
        Garage                  & 1.07 & [1.06, 1.08] \\
        Hot Press               & 1.01 & [1.00, 1.02] \\
        Open Plan               & 1.01 & [1.00, 1.02] \\
        Refurbished             & 1.04 & [1.03, 1.05] \\
        South Facing            & 1.02 & [1.01, 1.03] \\
        Ground Floor Apartment  & 0.97 & [0.95, 0.99] \\
        First Floor Apartment   & 0.98 & [0.96, 1.01] \\
        Second Floor Apartment  & 0.98 & [0.94, 1.02] \\
        Penthouse Apartment     & 1.14 & [1.08, 1.19] \\
        \hline
    \end{tabular}
    \caption{GAM 3 model coefficients of linear terms}
    \label{tab:mod_coeffs}
\end{table}

\begin{figure}[h!]
    \centering
    \includegraphics[width=\textwidth]{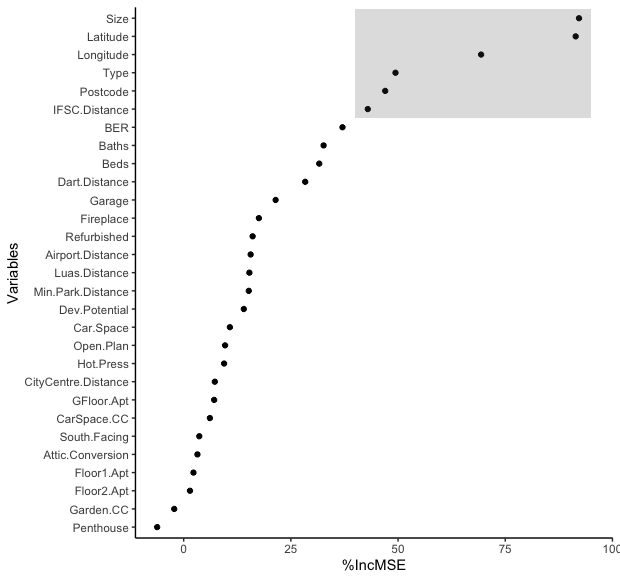}
    \caption{Increase associated with excluding each variable in percentage Mean Squared Error (MSE). Those highlighted by the grey box are some of the variables inherently included in our nearest neighbours approach. This plot showcases the most important variables in the random forest.}
    \label{fig:RandomForest_Plots}
\end{figure}

\begin{table}[h!]
    \centering
    \caption{Estimates and 95\% Confidence Interval, Corrected Postcode from GAM 5 model}
    \label{tab:CP_Tab1}
    \begin{tabular}{l c c} 
        \hline
        \textbf{Postcode} & \textbf{Estimate} & \textbf{ 95\% Confidence Interval} \\ 
        % & & \textbf{Interval} \\
        \hline
        Dublin 1            & 1.09 & [1.03, 1.16] \\
        Dublin 2            & 1.18 & [1.11, 1.25] \\
        Dublin 3            & 1.04 & [0.97, 1.11] \\
        Dublin 4            & 0.96 & [0.90, 1.02] \\
        Dublin 5            & 0.99 & [0.92, 1.07] \\
        Dublin 6            & 1.02 & [0.96, 1.08] \\
        Dublin 6W           & 1.00 & [0.94, 1.06] \\
        Dublin 7            & 1.17 & [1.11, 1.24] \\
        Dublin 8            & 1.10 & [1.05, 1.16] \\
        Dublin 9            & 1.13 & [1.06, 1.21] \\
        Dublin 10           & 0.85 & [0.77, 0.93] \\
        Dublin 11           & 1.03 & [0.95, 1.12] \\
        Dublin 12           & 0.95 & [0.89, 1.02] \\
        Dublin 13           & 1.01 & [0.91, 1.12] \\
        Dublin 14           & 0.95 & [0.89, 1.01] \\
        Dublin 15           & 1.15 & [1.06, 1.24] \\
        Dublin 16           & 1.00 & [0.93, 1.07] \\
        Dublin 17           & 1.04 & [0.94, 1.15] \\
        Dublin 18           & 0.95 & [0.87, 1.03] \\
        Dublin 20           & 1.01 & [0.92, 1.12] \\
        Dublin 22           & 0.83 & [0.75, 0.92] \\
        Dublin 24           & 0.86 & [0.79, 0.94] \\
        North County Dublin & 1.01 & [0.86, 1.18] \\
        South County Dublin & 0.88 & [0.81, 0.95] \\
        West County Dublin  & 0.94 & [0.90, 0.97] \\
        \hline
    \end{tabular}
\end{table}

%%%%%%%%%%%%%%%%%%%%%%%% CAME WITH TEMPLATE %%%%%%%%%%%%%%%%%%%%%%%%

%\begin{figure}
%  \includegraphics{example.eps}
%\end{figure}
%\begin{figure*}
%  \includegraphics[width=0.75\textwidth]{example.eps}
%\end{figure*}
% \begin{table}
% \caption{Please write your table caption here}
% \label{tab:1}       % Give a unique label
% \begin{tabular}{lll}
% \hline\noalign{\smallskip}
% first & second & third  \\
% \noalign{\smallskip}\hline\noalign{\smallskip}
% number & number & number \\
% number & number & number \\
% \noalign{\smallskip}\hline
% \end{tabular}
% \end{table}

% \begin{thebibliography}{}
% Author, Article title, Journal, Volume, page numbers (year)
% Author, Book title, page numbers. Publisher, place (year)
% etc
% \end{thebibliography}

\end{document}